\documentclass[runningheads]{llncs}
\usepackage[T1]{fontenc}

\usepackage{graphicx}
\AtBeginDocument{%
  }
\usepackage{amssymb}
\usepackage{amsmath}
\usepackage{mathtools}

\makeatletter
\renewcommand\subsubsection{\@startsection{subsubsection}{3}{\z@}%
                       {-8\p@ \@plus -4\p@ \@minus -4\p@}
                       {-0.5em \@plus -0.22em \@minus -0.1em}%
                       {\normalfont\normalsize\bfseries\boldmath}}
\makeatother
\usepackage{color}
\usepackage{url}
\usepackage{subcaption}
\usepackage{tabularray}
\usepackage{comment}
\usepackage{array} 
\usepackage{graphicx} 
\usepackage{algorithm}
\usepackage{algorithmic}
\usepackage{stfloats}
\usepackage{booktabs}
\usepackage{placeins}
\usepackage{appendix}
\usepackage{float}

\begin{document}
\title{XION: Interpreting Network Traffic
Anomalies via Dependency Violations}
\titlerunning{Interpreting Network Traffic
Anomalies via Dependency Violations}

\author{Federica Uccello\inst{1}\orcidID{0000-0001-9243-7047} \and
Simin Nadjm-Tehrani\inst{1}\orcidID{0000-0002-1485-0802}}
\authorrunning{Uccello et al.}
%
\institute{Department of Computer and Information Science, Linköping University, Sweden 
\email{federica.uccello@liu.se}\\\email{simin.nadjm-tehrani@liu.se}\\
}
\maketitle              
\begin{abstract}
Current research on security monitoring is increasingly focusing on machine-learning-based approaches, but caveats remain. 
In addition to huge computational overhead, one concern is the lack of insights into ``why" alerts are raised. Existing interpretability approaches rely on feature attribution methods that ignore dependencies among features or on causal modeling that requires extensive domain knowledge or computational resources.
This work proposes XION, a method for modeling relationships among network-flow features based on benign traffic only. During detection, anomalies are identified through violations of expected feature dependencies. Further, XION supports post-alert analysis by identifying which feature relationships break, when they break along the attack timeline, and how dependency violations evolve relatively to other identified violations.
XION is evaluated on standard IDS datasets and compared against an Isolation Forest (IF) baseline across multiple attack scenarios, including both volumetric and stealthier attacks. Results show that XION matches or exceeds IF recall in all evaluated scenarios, while requiring up to 7$\times$ less inference time.
At the post-alert stage, the dependency violation analysis reveals temporal and structural patterns consistent with known attack behaviors, and is further evaluated against SHAP-based feature attribution of IF alerts. 
Together, these findings confirm that attacks indeed disrupt feature dependencies learned from benign traffic, and that these disruptions provide additional information for understanding an alert.
\end{abstract}

\section{Introduction}
In the complex landscape of network security, Intrusion Detection Systems (IDSs) remain a fundamental tool for identifying malicious or abnormal network activity. To overcome the limitations of traditional signature-based IDSs, research has increasingly shifted toward anomaly-based approaches that leverage Machine Learning (ML) techniques to learn patterns of benign network behavior and raise alerts when deviations are identified \cite{apruzzese2023role}.

Despite promising results in theory and test environments, significant challenges remain when applying ML to intrusion detection 
\cite{viegas2023toward,apruzzese2023role,arp2024pitfalls}.
Many ML-based detectors can introduce expensive computational overhead \cite{bolon2024review} and suffer from limited interpretability. Although they may successfully detect anomalous behavior, there is little insight into \textit{why} an alert was triggered or how security operators should act upon it. 

"Explainability" techniques have attempted to tackle the interpretability problem by identifying features that contribute to ML models' outputs \cite{hassija2024interpreting,saeed2023explainable}. However, state-of-the-art approaches exhibit limitations \cite{ademi2025pomelo,marques2024logic,uccello2026feature,huang2024failings}, and they typically treat features as independent variables, while network traffic characteristics are interdependent and evolve through complex relationships. 

In this work, we explore whether learning feature dependencies under benign conditions can provide insights into anomalies observed during the operation of a network.

Our intuition is that if a feature breaks its typical relationships it could be related to a structural disruption in the system. We look at indicators closely associated with the violations at inference time to provide hints for diagnosis. 
Analyzing \emph{which} relationships break and \emph{when} provides a structured view of anomalous behavior that is not available from aggregate anomaly scores alone or standard explainability techniques. It turns out that this process is also efficient in using resources.

The contributions of this work are as follows:
\begin{itemize}
    \item We propose a novel method (XION) based on modeling feature-level dependencies and use their violations as a unified signal for both anomaly detection and post-alert analysis.
    \item We show that the learned feature dependencies are violated during attack scenarios, producing anomaly signals as good as an Isolation Forest (IF) baseline in two open IDS datasets, but with much lower inference time.
    \item We demonstrate that dependency violation analysis can help prioritize features for post-alert investigation and reveal attack-specific patterns over time, without requiring a separate post-hoc explanation step.
\end{itemize}

The remainder of the paper is organized into the following sections: Section \ref{sec:related} summarizes related work and positions the study within the state-of-the-art. Section \ref{sec:background} introduces the terminology and theoretical concepts needed to understand XION, which is described in Section \ref{sec:method}, along with the threat model and research questions. Section \ref{sec:experiments} describes the data and pre-processing, the experimental approach, and the evaluation strategy. The results are presented and discussed in Section \ref{sec:results}. Section \ref{sec:lim_fw} discusses threats to validity and future directions. Finally, Section \ref{sec:conclusion} concludes the paper.

\section{Related Work}
In this section, we restrict the comparisons to methods that model dependencies with some transparency within anomaly detection. Works addressing attack detection using heavy-duty ML are outside our scope.
\label{sec:related}
\subsection{Anomaly Detection}
We review here the related works that are based on observations of dependencies in the domain features.
\subsubsection{System-level dependencies}
Provenance-based approaches capture structural and temporal relationships among interacting entities. Yang et al. \cite{yang2023prographer} propose a provenance-based anomaly detection system that models structural and temporal relationships among system entities through provenance graph embeddings.
Similarly, Li et al. \cite{li2025exploring} leverage provenance graphs and Answer Set Programming to express attack patterns as logical rules over system interactions.
Such approaches provide rich semantic representations at the cost of computational overhead and expert effort. XION instead models statistical dependencies among aggregate network-flow features, with low training and inference time.

\subsubsection{Traffic-feature dependencies}
Closer to our setting, existing approaches exploit statistical relationships within network traffic. Mirsky et al. \cite{mirsky2018kitsune} group correlated traffic features and model these groups using an ensemble of autoencoders, detecting anomalies through reconstruction errors.
Lin and Nadjm-Tehrani \cite{lin2023protocol} study server-driven traffic in Supervisory Control And Data Acquisition (SCADA) networks by modeling inter-flow dependencies and detecting anomalies through deviations in the covariance structure of multivariate flow time series learned under benign conditions. 
Both demonstrate that relationships within benign traffic can provide useful anomaly signals. XION instead explicitly models dependencies between individual flow features, allowing specific relationship violations to be identified and subsequently analyzed, using the same information for detection and to support post-alert analysis.

\subsubsection{Graphical Models and Generic Anomaly Detection}
From a modeling perspective, XION is closely related to graphical methods for representing conditional dependencies. 
Hallac et al. \cite{hallac2017network} propose Time-Varying Graphical Lasso (TVGL), which estimates a sequence of sparse inverse covariance matrices to track changes in conditional dependencies over time. This is a generic method, not developed for cybersecurity applications. While both methods use sparse precision matrices to represent dependencies, TVGL repeatedly estimates the dependency structure and analyzes how the graph itself changes over time, requiring a dedicated optimization procedure for scalable inference. XION learns a single dependency structure from benign traffic and uses it to define local feature-prediction models. TVGL addresses a different problem and is not a direct anomaly-detection baseline, but our approach uses the same mathematical concept in a cybersecurity context.


Another related approach is DepAD, a general anomaly detection framework proposed by Lu et al. \cite{lu2025depad}, which uses feature selection and deviations from expected feature values to detect anomalies. While conceptually similar, XION targets higher-dimensional network traffic data and additionally examines the temporal ordering of
dependency violations for post-alert analysis.

IF 
has demonstrated strong detection capability and scalability in high-dimensional anomaly detection tasks, including competitive performance against other unsupervised models \cite{liu2012isolation}, and is well-suited for tabular data, where tree-based ensembles often outperform deep learning approaches \cite{shwartz2022tabular}. While deep neural networks and, more recently, large language models have also been explored for network intrusion detection \cite{zoppi2024anomaly,ferrag2025generative}, they do not necessarily improve performance on tabular network data and can entail substantially higher inference and computational costs \cite{bolon2024review,uccello2026flies}. 
In our evaluation, we use IF as a baseline. 

\subsection{Post-Alert Analysis}

\subsubsection{Feature Attribution}
A line of research defines root causes as features with high importance in anomaly detection models. They do not model causality or dependencies explicitly, but treat important features as a proxy for cause.

Carletti et al. \cite{carletti2019explainable} decompose IF anomaly scores to rank features by their contribution to the isolation process, while Roelofs et al. \cite{roelofs2021autoencoder} identify features contributing most to autoencoder anomaly scores in wind turbine SCADA data. Kuk et al. \cite{kuk2023feature} analyze SHAP-based feature importance \cite{lundberg2017shap} over time to identify early indicators of asset degradation.  Inspired by
their temporal analysis, we examine how SHAP feature attributions evolve
over the attack timeline and use this as a baseline for comparison with XION's post-alert analysis.

While computationally lightweight, these approaches implicitly equate feature attribution with causation. In comparison, we explicitly model feature dependencies and investigate the temporal progression of dependency violations across related features.

\subsubsection{Causality Modeling}
Other works investigate post-alert analysis through explicit causal reasoning. 
Tonon et al. \cite{tonon2025radice} and Wang et al. \cite{wang2023interdependent} perform root cause localization by modeling causal dependencies among system metrics through causal graphs and graph neural networks, respectively. 

Lin et al. \cite{lin2025claric} propose 
a framework for root cause analysis in smart manufacturing through causal and counterfactual tests that evaluate whether correcting a factor (feature or feature group) reduces anomalous behavior.
These approaches require causal modeling and, in some cases, extensive domain knowledge. XION instead provides a lightweight alternative based on statistical dependencies learned directly from benign traffic, exposing which relationships are disrupted and how, without making causal claims.

\subsubsection{Dependency-Aware Interpretation.}
Certain works model dependencies among features to improve the interpretability of ML models. Aas et al. \cite{aas2021explaining} address the feature independence assumption of SHAP, the de facto standard for feature attribution.
They estimate conditional feature distributions using Gaussian, copula-based, or empirical approaches. 
With a similar motivation, Salih \cite{salih2024explainable} proposes a global feature attribution method that models how features can be used to predict one another through a series of univariate regressions. 
Beyond first-order feature attribution, Anthony et al. \cite{anthony2025rule} 
introduce an interaction-aware extension of SHAP that quantifies pairwise feature interactions, capturing nonlinear and conditional dependencies between binary indicators. 
In network anomaly detection, Shafir et al. \cite{shafir2025explainable} combine normalizing flows with Shapley values to identify discriminative traffic features and explain anomalous instances.

Unlike these works, XION treats violations of learned dependencies as the anomaly signal itself.
The dependency structure is learned solely from benign traffic, without requiring malicious samples. Moreover, XION preserves individual dependency violations, allowing us to examine \emph{which} relationships break and \emph{how} these violations evolve over time during attacks.

Taken together, these distinctions position XION as a middle ground between aggregate anomaly scores and full causal discovery, offering additional information compared to classical anomaly detectors at low computational cost.

\section{Background}
\label{sec:background}
The concepts underlying XION include statistical dependencies, their distinction from correlation and causality, and their representation through graphical models.

\subsection{Correlation, Causality, and Dependency}
Reasoning about modern complex systems through high-fidelity models is a daunting task, if not impossible. In practice, system behavior is typically assessed through observed variables (features), while the underlying processes are often treated as a black box. As a result, the problem reduces to understanding patterns and relationships within observed feature values over time.

Correlation, dependency, and causality are related but distinct concepts that are often conflated. Correlation measures the statistical association between variables, quantifying the extent to which they vary together \cite{altman2015points,morin2003correlation}. Correlated variables may exhibit similar behavior, but correlation alone does not imply that one variable influences the other.

Causality instead refers to directional relationships in which changes in one variable directly affect another variable. Unlike correlation, causal relationships imply intervention semantics: modifying a cause can alter its effects \cite{pearl2009causality}. 

Dependency describes whether the value of one variable provides information about another. Two variables are statistically independent if knowledge of one does not change the probability distribution of the other. Conditional dependency extends this concept by considering relationships between variables after accounting for additional variables. Conversely, two variables are conditionally independent if they become independent once the conditioning variables are known \cite{koller2009probabilistic}.

Dependency modeling can be viewed as an intermediate level of analysis between correlation and full causal inference. Unlike correlation, conditional dependencies capture direct statistical relationships while accounting for the influence of other variables. At the same time, dependency modeling does not require the strong assumptions, interventions, or domain knowledge typically needed for causal discovery. 

\subsection{Gaussian Graphical Models}
Gaussian graphical models (GGMs) provide a framework for modeling conditional dependencies among random variables \cite{friedman2008sparse}. Let $X \in \mathbb{R}^d$ denote a random vector with covariance matrix $\Sigma$ and precision matrix $\Theta = \Sigma^{-1}$. The covariance matrix captures marginal correlations between variables, whereas the precision matrix encodes conditional dependencies under Gaussian assumptions.

Let $X_i$ denote the $i$-th random variable. In Gaussian graphical models, zeros in the precision matrix correspond to conditional independence after accounting for all remaining variables:

\begin{equation}
\label{eq:ind}
{\Theta}_{ij} = 0
\quad \Longleftrightarrow \quad
X_i \perp X_j \mid X_{-\{i,j\}} ,
\end{equation}

where $X_{-\{i,j\}}$ denotes the collection of all variables except $X_i$ and $X_j$, $\perp$ denotes statistical independence, and $\mid$ denotes conditioning on the variables to its right.

If ${\Theta}_{ij} \neq 0$, variables  $X_i$ and  $X_j$ are conditionally dependent given the remaining variables.
$\Theta$ can be interpreted as a graph, where nodes correspond to variables and an edge is placed between nodes $i$ and $j$ whenever $\Theta_{ij} \neq 0$. Since $\Theta$ is symmetric, the resulting graph is undirected. In this representation, an edge indicates that variables $X_i$ and $X_j$ are conditionally dependent given all remaining variables, whereas the absence of an edge ($\Theta_{ij} = 0$) corresponds to conditional independence.

GGMs assume that the observed variables are jointly Gaussian. In practice, data distribution may violate this constraint. The nonparanormal model from Liu et al. \cite{liu2009nonparanormal} extends GGMs to non-Gaussian data, applying monotonic transformations to each variable such that the transformed data is approximately Gaussian. This is implemented using rank-based or quantile transformations.

\section{Proposed Method}
\label{sec:method}

Our work is based on the idea that the features observed in network telemetry do not behave independently. Instead, they exhibit inter-dependencies that reflect normal system behavior. 
When attacks occur, these relationships may be disrupted as the system transitions into abnormal states. Hence, we believe dependency violations provide signals for interpreting anomaly alerts by prioritizing features for post-alert analysis.

We therefore build a workflow/proof of concept and perform experiments to investigate the following research questions:

\begin{itemize}
    \item \textbf{RQ1:} Is there evidence that dependency violations occur during attack instances, and how do these signals relate to alerts produced by an ML-based anomaly detector?
    \item \textbf{RQ2:} Can dependency violation signals be used to prioritize features for post-alert analysis, and do these prioritized features align with domain knowledge about attack behavior?
    \item \textbf{RQ3:} How time-efficient is dependency-based anomaly detection and post-alert analysis compared to an ML-based detector and post-hoc feature attribution?
\end{itemize}

\begin{figure*}
    \centering
    \includegraphics[width=0.85\linewidth]{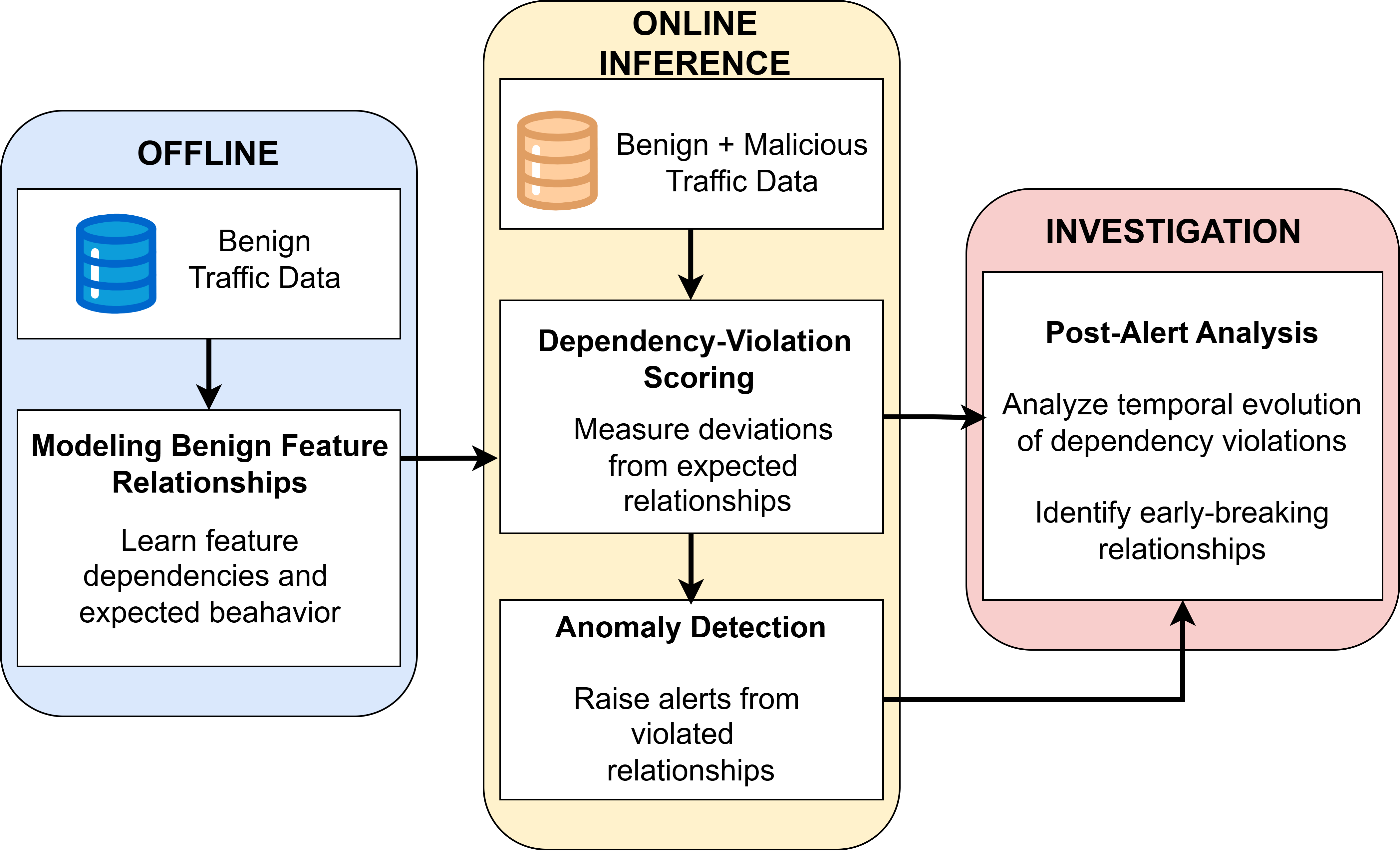}
    \caption{High-level overview of XION.}
    \label{fig:method}
\end{figure*}

The proposed method, XION, learns feature dependencies from benign network traffic and analyzes their violations during anomalous activity.
Figure \ref{fig:method} shows a high-level view of XION. Here, the blue blocks refer to offline operations, the yellow ones to inference time, and the red ones to post-alert investigation. 

To quantify abnormal behavior at alert time, we first model how features relate to one another under benign conditions (Algorithm \ref{alg:model}, explained in Section \ref{sub:alg1}). 
The dependency modeling follows theory from Friedman et al. \cite{friedman2008sparse}, where random variables correspond to features. Following the nonparanormal model from Liu et al. \cite{liu2009nonparanormal}, we apply a quantile transformation to map each feature to an approximately Gaussian distribution. The transformation is learned on benign data and consistently applied when computing dependency violations (Algorithm \ref{alg:violation}, explained in Section \ref{sub:alg2}). The violations are used as a unified signal for anomaly detection (Section \ref{sub:method_ad}) and post-alert analysis (Algorithm \ref{alg:pa}, explained in Section \ref{sub:pa_method}). 

\subsection{Threat Model}
\label{sub:threat_model}
We consider an adversary whose goal is to disrupt service availability, obtain unauthorized access, exploit vulnerable services, or gather information about the monitored system or network. The adversary can interact with exposed hosts and services and generate malicious network activity that alters one or more observable flow-level characteristics.

The defender passively observes network-flow features and learns the expected dependency structure from benign traffic collected prior to deployment. At inference time, the detector evaluates new traffic against these learned relationships. The attacker is assumed to have no access to, or control over, the benign training data, learned dependency graph, model parameters, or detection thresholds.

Our threat model concerns attacks against the monitored system. Data poisoning, manipulation of the monitoring infrastructure, and adaptive evasion designed with knowledge of the detector are outside the scope of this work.

\subsection{Modeling Benign Feature Relationships}
\label{sub:alg1}

In the offline phase of XION, given a benign training dataset $X \in \mathbb{R}^{n \times d}$ with $n$ samples and $d$ features, we estimate a sparse precision matrix ${\Theta}$ using \textsc{GraphicalLassoCV}\footnote{\url{https://scikit-learn.org/stable/modules/generated/sklearn.covariance.GraphicalLassoCV.html}} (lines 1--2 of Algorithm \ref{alg:model}). 

From the sparsity pattern of ${\Theta}$, we define for each feature $i$ the neighbor set ${N}(i)$. This induces an undirected dependency graph representing the normal structural relationships among features (line 5), with features corresponding to nodes and edges indicating conditional dependencies between neighboring features.

The dependency graph provides structural information about which features are directly related, but not how their values depend on one another. Intuitively, if features are conditionally dependent, the value of one feature should be predictable from those of its neighbors. To model these relationships, we train local predictive models as described below.

If $\mathcal{N}(i) = \emptyset$, no predictive model is trained and the feature is treated as independent in the learned graph (lines 6--8). 

For each feature $i$ with a non-empty neighbor set, we train a ridge regressor \cite{hastie2009elements} $f_i$ using benign data to predict the value of feature $i$ from its neighbors (lines 10--12).

\begin{equation}
\label{eq:ridge}
f_i : \mathbb{R}^{|\mathcal{N}(i)|} \rightarrow \mathbb{R}.
\end{equation}

The regressor models the expected values of feature $i$ from the neighboring features under benign conditions. Then, residuals are  computed across the benign training data as the difference between observed and predicted values :

\begin{equation}
\label{eq:res}
\mathcal{R}_i = X_{i} - X_{i}^{N},
\end{equation}

where $X_i^{N}$ denotes the vector of predicted values $f_i(x_{\mathcal{N}(i)})$, $x_{\mathcal{N}(i)} \in \mathbb{R}^{|\mathcal{N}(i)|}$ the vector of neighboring feature values, and $\mathcal{R}_i$ the vector of residuals. The residual scale, defined as $\sigma_i = \mathrm{Std}(\mathcal{R}_i)$, captures the typical benign prediction error for feature $i$.

After this stage, for each feature $i$ we retain its neighbor set $\mathcal{N}(i)$, local regressor $f_i$ (if defined), and benign residual scale $\sigma_i$ (line 15). These elements will later be used to quantify relationship violations at inference time (Algorithm \ref{alg:violation}).

\begin{algorithm}[t]
\caption{Modeling Feature Relationships under Benign Conditions}
\label{alg:model}
\begin{algorithmic}[1]
\REQUIRE Benign dataset $X \in \mathbb{R}^{n\times d}$
\ENSURE Neighbors $\{\mathcal{N}(i)\}$, local regressors $\{f_i\}$, residual scales $\{\sigma_i\}$

\STATE Fit \textsc{GraphicalLassoCV} on $X$ to estimate a sparse precision matrix ${\Theta}$
\STATE Set ${\Theta}_{ii} \leftarrow 0$ for all $i$

\FOR{$i=1$ \TO $d$}
  \STATE \# Define neighbors:
  \STATE \quad $\mathcal{N}(i) \leftarrow \{ j \neq i : {\Theta}_{ij} \neq 0 \}$
  
  \IF{$\mathcal{N}(i)=\emptyset$}
     \STATE $f_i \leftarrow \texttt{None}$
     \STATE $\sigma_i \leftarrow \textsc{Std}(X_{i})$
  \ELSE
     \STATE Train ridge regressor $f_i$ to predict $X_{i}$ from neighbors
     \STATE $X^N_i \leftarrow f_i(x_{\mathcal{N}(i)})$
    \STATE $\sigma_i \leftarrow \textsc{Std}(X_i - X_i^{N})$
  \ENDIF
\ENDFOR

\RETURN $\{\mathcal{N}(i)\}, \{f_i\}, \{\sigma_i\}$
\end{algorithmic}
\end{algorithm}

\subsection{Dependency Violation Scoring}
\label{sub:alg2}

At inference time, given an instance $x \in \mathbb{R}^d$ 
(a sample to be evaluated), we quantify how strongly each feature deviates from its expected relationship with its neighbors under benign conditions (Algorithm \ref{alg:violation}).

If no regressor is available for feature $i$,  we set $v_i = 0$, as no relationship-based violation can be evaluated (lines 1--5 of Algorithm \ref{alg:violation}). If a local regressor $f_i$ was learned during the benign modeling stage, we use the regressor to predict the value of feature $i$ from its neighbors, and define the normalized residual magnitude (violation score) as (lines 6--7):

\begin{equation}
\label{eq:violation}
v_i =
\frac{
\left| x_i - \hat{x}_i \right|
}{
\sigma_i
},
\end{equation}

where $x_i$ is the observed value of feature $i$ in the input instance, $\hat{x}_i$ is the value predicted from the neighboring features, and $\sigma_i$ is the benign residual scale estimated during the offline phase.

The resulting vector $v = (v_1, \dots, v_d)$ quantifies, for each feature with neighbors, how strongly its observed value deviates from that expected under the conditional relationships learned from benign data  (line 9). 

\begin{algorithm}[t]
\caption{Per-feature Dependency Violation Scoring}
\label{alg:violation}
\begin{algorithmic}[1]
\REQUIRE Instance $x \in \mathbb{R}^{d}$, neighbors $\{\mathcal{N}(i)\}$, local regressors $\{f_i\}$, benign residual scales $\{\sigma_i\}$
\ENSURE Violation score vector $v = (v_1, \dots, v_d)$

\STATE Initialize $v_i \leftarrow 0$ for all $i \in \{1,\dots,d\}$

\FOR{$i=1$ \TO $d$}
  \IF{$f_i = \texttt{None}$ \OR $\mathcal{N}(i)=\emptyset$}
     \STATE \textbf{continue}
  \ENDIF
  \STATE $\hat{x}_i \leftarrow f_i\!\left(x_{\mathcal{N}(i)}\right)$
  \STATE $v_i \leftarrow 
  \dfrac{\left| x_i - \hat{x}_i \right|}{\sigma_i}$
\ENDFOR

\RETURN $v = (v_1, \dots, v_d)$
\end{algorithmic}
\end{algorithm}

\subsection{Anomaly Detection}
\label{sub:method_ad}
Dependency violation vectors are used to discriminate between benign and anomalous traffic. A relationship is considered broken when the corresponding violation score exceeds a threshold $\tau$. 

Given an instance $x \in \mathbb{R}^d$, the number of violated relationships is computed as:

\begin{equation}
S(x) =
\left|
\left\{
i \in \{1,\dots,d\}
\;:\;
v_i > \tau
\right\}
\right|.
\label{eq:alert}
\end{equation}

An instance is deemed anomalous if 
$S(x) \geq k$, where $k$ is the minimum number of simultaneous dependency violations required to raise an alert. In this work, we set $k=2$, requiring at least two simultaneous dependency violations to raise an alert and reducing the likelihood of spurious detections caused by isolated fluctuations.
$\tau$ was derived from a held-out benign validation set. In the rest of the paper, the component derived using this process is referred to as the ``Dependency-based detector" (DP detector).

\subsection{Post-Alert Analysis}
\label{sub:pa_method}
In the investigation phase of XION, dependency violations are analyzed to determine whether they provide signals for interpreting anomalous behavior. In particular, the temporal order in which feature relationships break during attacks is investigated (Algorithm \ref{alg:pa}). 

\begin{algorithm}[t]
\caption{Post-Alert Dependency Analysis}
\label{alg:pa}
\begin{algorithmic}[1]
\REQUIRE Window $W$, violation scores $\{v_i^t\}_{t\in W}$, thresholds $\tau,k$, neighbors $\{\mathcal{N}(i)\}$
\ENSURE First violations $\{t_i^*\}$,
neighbor sets and counts

\FOR{$i=1$ \TO $d$}
    \STATE $\mathcal{T} \leftarrow
    \{t\in W:S(x^t)\ge k \land v_i^t>\tau\}$ \# $S$ defined in Eq.~\eqref{eq:alert}
    \STATE $t_i^* \leftarrow
    \begin{cases}
        \bot, & \mathcal{T}=\emptyset,\\
        \min(\mathcal{T}), & \text{otherwise}.
    \end{cases}$
\ENDFOR

\FOR{$i=1$ \TO $d$}
    \IF{$t_i^*\neq\bot$}
        \STATE $\mathcal{N}_i^* \leftarrow
        \{j\in\mathcal{N}(i):t_j^*\neq\bot\}$
        \STATE $\mathcal{B}_i \leftarrow
        \{j\in\mathcal{N}_i^*:t_j^*<t_i^*\}$
        \STATE $\mathcal{C}_i \leftarrow
        \{j\in\mathcal{N}_i^*:t_j^*=t_i^*\}$
        \STATE $\mathcal{L}_i \leftarrow
        \{j\in\mathcal{N}_i^*:t_j^*>t_i^*\}$
        \STATE $(B_i,C_i,L_i) \leftarrow
        (|\mathcal{B}_i|,|\mathcal{C}_i|,|\mathcal{L}_i|)$
    \ENDIF
\ENDFOR

\RETURN $\{t_i^*\},(\mathcal{B}_i,\mathcal{C}_i,\mathcal{L}_i)$
\end{algorithmic}
\end{algorithm}

Post-alert analysis is performed within an observation window $W$ and considers only instances that trigger DP alerts. For each feature $i$, its first violation $t_i^*$ among these instances within the window W is identified (lines 1--4 of Algorithm \ref{alg:pa}). If no violation is observed, $t_i^*=\bot$, where $\bot$ denotes no violations in this window.

For each feature with a defined first violation, $t_i^*$ is compared with those of its neighbors in $\mathcal{N}(i)$ that also have defined first violations (lines 5--11). The sets
$\mathcal{B}_i$, $\mathcal{C}_i$, and $\mathcal{L}_i$ identify the neighboring features whose first violations occur before, at the same time as, or after that of feature $i$, respectively.
Their cardinalities, $B_i$, $C_i$, and $L_i$, summarize these temporal relationships. The intuition is that features whose relationships break earlier than those of their neighbors may be associated with early attack indicators, whereas features that break later may reflect how the anomaly manifests across the system over time.

The temporal progression of dependency violations is further characterized using three stages: Early Stage (ES), Mid Stage (MS), and Late Stage (LS). Let $t_s$ and $t_e$ denote the time of the start of an alert sequence and the time of the last alert in that sequence, respectively, both within the observation window W ( $t_ s < t_e$,). Then, using the timestamps
of the first and last DP alerts observed, the relative position of each feature's first violation is defined as:

\begin{equation}
r_i = \frac{t_i^* - t_s}{t_e - t_s}.
\end{equation}

Each feature with a defined first violation is
assigned to a temporal stage according to:

\begin{equation}
\operatorname{Stage}(i) =
\begin{cases}
\mathrm{ES}, & r_i \leq \alpha,\\
\mathrm{MS}, & \alpha < r_i < 1-\beta,\\
\mathrm{LS}, & r_i \geq 1-\beta,
\end{cases}
\end{equation}

where $\alpha$ and $\beta$ specify the early-
and late-stage fractions of the alert interval,
respectively.

\section{Experiments and Evaluation Strategy}
\label{sec:experiments}

The evaluation addresses the RQs introduced in Section \ref{sec:method}. 
First, the distributions of dependency violation scores during benign and attack instances are examined to observe whether attack periods can be distinguished from normal periods, and the resulting anomaly detection performance is compared with that of an ML-based detector (RQ1). Next, the usefulness of dependency violations for prioritizing features in post-alert analysis is assessed by examining their temporal patterns and evaluating whether the prioritized features align with domain knowledge (RQ2). Finally, execution times are measured to compare dependency-based detection and
post-alert analysis with the baselines (RQ3)\footnote{The source code and data used in this study are available on GitHub{\url{https://github.com/FedeU95/XION/}. Will be made available on request and public upon paper acceptance.}. All the experiments were executed in a Debian-based virtual machine equipped with an Intel Core Ultra 7 155U CPU and 32 GB RAM.}.

\subsection{Datasets and Pre-Processing}
\label{sub:data}
XION is tested on two different IDS datasets: CIC-IDS-2018 \cite{Sharafaldin2018TowardGA}, and CIC-UNSW-NB15 \cite{mohammadian2024data} to cover both a diverse set of attacks and show that XION is not tailored to a given dataset. 

CIC-IDS-2018 is a well-established benchmark for intrusion and anomaly detection. The dataset contains multiple traffic flows collected over different days, and includes both benign data and different attack scenarios. To encompass diverse adversarial behavior, the evaluation includes HTTP flood Distributed Denial of Service (DDoS), Slowloris, and SSH brute-force attacks.

CIC-UNSW-NB15 is an intrusion detection dataset selected to test the dependency modeling with a higher number of available benign samples. Based on behavioral diversity and sufficient representation in the dataset, the attacks for evaluation include DoS, Exploit, and Reconnaissance. 

All data is aligned to a common feature schema derived during pre-processing.  
Both datasets have the same features, and details on feature semantics can be found in the CIC-IDS-2018 documentation \cite{Sharafaldin2018TowardGA}.
 
In our tests, each benign set is partitioned into three roles: training, validation, and test. 
The training split is used to learn the dependency model and fit the ML baseline, while the validation split is used for the configuration of the detectors. The final evaluation is then performed on the benign test split together with all available attack samples.
Since the evaluation is unsupervised and part of our goal is to analyze the violation of feature dependencies over time, the test sets are not artificially balanced or subsampled. Preserving the original sample distributions also allows temporal effects to be analyzed without introducing selection bias.

See Appendix \ref{app:dataset} for a detailed description of each attack, the pre-processing steps, and the train/test/evaluation splits.

\subsection{Dependency Graph under Benign Conditions}
Dependencies between features are modeled following Algorithm \ref{alg:model} (Section \ref{sub:alg1}). 

The coefficient of determination ($R^2$) is computed to evaluate the quality of the learned feature dependencies under benign conditions. $R^2$ ranges from 0 to 1 and measures the goodness-of-fit of the local predictive models, with values closer to 1 indicating more accurate predictions from neighboring features under benign conditions. In our experiments, $R^2$ is computed both on training and validation data to evaluate whether the model generalizes to unseen benign instances. 

We also inspect the learned dependency graph to assess whether it captures meaningful relationships beyond trivial associations between closely related features.

\subsection{Dependency Violation as Anomaly Signals}
\label{sub:eval_rq1}
To answer RQ1, the distributions of dependency violations under benign and attack conditions are compared. Then, their relationship with alerts produced by an ML-based anomaly detector is examined through detection metrics and temporal overlap analysis.

\subsubsection{Dependency Violation Distribution Analysis}
\label{sub:viol}
To assess whether dependency violation behavior differs between benign and attack instances, the distributions of violation scores (Algorithm \ref{alg:violation}) are compared using two statistical tests: the two-sided Mann--Whitney U test~\cite{macfarland2016mann}, to assess whether violation scores tend to be systematically larger under attack conditions, and the Kolmogorov--Smirnov (KS) test \cite{berger2014kolmogorov}, to detect broader distributional differences including changes in variance and tail behavior. Previous work shows that these tests exhibit different power characteristics depending on the underlying distribution \cite{ozccomak2013comparison}, hence their combined use allows a more inclusive assessment of distributional differences. 

The Mann--Whitney U test returns a $p$-value indicating whether the benign and attack distributions differ significantly. Small $p$-values ($p < 0.05$) suggest that the observed differences are unlikely to arise by chance under the null hypothesis that both samples originate from the same distribution. 

The KS test additionally provides the KS statistic, which measures the maximum distance between the empirical cumulative distributions of benign and attack samples and ranges from 0 to 1. While the $p$-value indicates whether a significant difference exists, the KS statistic quantifies the magnitude of distributional separation. Higher KS values correspond to stronger divergence between benign and attack traffic.

The analysis is performed on two flow-level metrics: (i) the mean violation score $v_{mean}$ across features, which captures the overall magnitude of dependency deviations within a flow, and (ii) the median number of violated features per flow $n_{viol}$, which quantifies how many learned feature relationships simultaneously deviate beyond the violation threshold $\tau$.

\subsubsection{Anomaly Detection and Baseline}
\label{sub:ad_exp}
The statistical analysis above is complemented by comparing the anomaly signals of the DP detector with those of an independently trained ML detector on the same datasets. Note that the aim is to assess whether dependency violations provide meaningful anomaly signals and how these relate to those produced by an established ML baseline, rather than to optimize detection performance.

The threshold $\tau$ (Section \ref{sub:method_ad}) was calibrated using the held-out benign validation set, yielding values of 2.5 for CIC-IDS-2018 and 3.0 for CIC-UNSW-NB15. These correspond to events expected to occur in approximately 1.24\% and 0.27\% of benign observations under a Gaussian assumption. 

Since our goal is a lightweight approach at inference time, we use IF as an ML baseline. For the IF baseline, each sample is assigned an anomaly score, computed as the negative IF sample score so that higher values correspond to more anomalous observations\footnote{See the Scikit-learn IF documentation for the sample score definition: \url{https://scikit-learn.org/stable/modules/generated/sklearn.ensemble.IsolationForest.html}.}. The alert threshold is calibrated on the benign validation set using a percentile-based strategy. The final operating threshold was selected as the 90th percentile of the benign anomaly-score distribution to balance false positive (FP) rates and detection performance across attack categories. 

\subsubsection{Evaluation Metrics for Anomaly Detection}
For both detectors, we measure Recall (Rec.), Balanced Accuracy (BA), Matthews correlation coefficient (MCC), false positive rate (FPR), and Specificity (Spec.) as sanity-check detection metrics. 
BA and MCC are selected to give an accurate evaluation of accuracy and overall detection quality, taking into account class imbalance. 
Specificity measures the proportion of true negatives correctly identified. We additionally evaluate ROC and precision--recall curves, together with AUROC and AUPRC, to evaluate detector performance across different decision thresholds\footnote{See Scikit-learn documentation for metrics definition: \url{https://scikit-learn.org/stable/api/sklearn.metrics.html}}.

\subsubsection{Temporal Overlap}
We also analyze the temporal overlap between the two anomaly signals. Specifically, we compute the IF anomaly score and the maximum dependency violation score over time. The maximum violation is used to capture the strongest deviation from learned feature relationships relative to the actual attack timeline. This analysis allows us to examine how the two detectors respond given an attack timeline.

\subsection{Dependency Violation for Post-Alert Analysis}
\label{sub:eval_rq2}
To answer RQ2, feature-level dependency violations are analyzed for each attack using the post-alert analysis procedure described in
Section \ref{sub:pa_method}. The resulting feature prioritization is then evaluated against literature-based references of known attack behavior.

\subsubsection{Post-Alert Analysis and Baseline}
Following Section \ref{sub:pa_method}, the first-violation ordering of neighboring features is examined, retaining their identities and relative order. In our experiments, the observation window $W$ is defined as a 24-hour period, with one day considered per attack. This choice is based on the open datasets we had available for evaluation. In a live network, the choice is dependent on some expectation (assumption) on the number of distinct ongoing attacks per period of time. 

The ES and LS thresholds $\alpha$ and $\beta$ are set to 5\% and 10\% of the reference interval, respectively. These thresholds are intended to distinguish violations emerging shortly after an attack onset from those appearing near its end.

As a baseline for post-alert interpretation, we use SHAP \cite{lundberg2017shap} to explain the alerts produced by the IF detector\footnote{We use the TreeSHAP optimization to provide an efficient implementation of SHAP for the tree-based IF model.}.  Inspired by the temporal feature-importance analysis of Kuk et al.~\cite{kuk2023feature}, we examine how SHAP feature attributions evolve over an attack timeline. 

To ensure a consistent comparison with the baseline, we use ground-truth attack boundaries to define the temporal stages (ES, MS, and LS), since the DP and IF detectors may produce different alerts. As a sanity check, XION's stage assignments are recomputed using the first and last DP alerts in $W$ as boundaries and compared with the ground-truth-based assignments (see Section \ref{sub:pa}).


\subsubsection{Evaluation Approach for Post-Alert Analysis}
To evaluate the post-alert evidence, we define a literature-grounded behavioral reference for each attack scenario. The reference summarizes the attack behavior documented by the dataset descriptions and related literature (see Appendix~\ref{app:literature}). It serves as an independent reference to assess whether the evidence exposed during post-alert analysis is consistent with known attack behavior. 

We consider both the features represented by the identified features and their temporal organization. Evidence is considered \textbf{supported} when the dominant feature and/or temporal evidence corresponds closely to the documented attack behavior, \textbf{partially supported} when this correspondence is indirect, non-specific, or provides limited temporal differentiation, and \textbf{not supported} when no clear correspondence can be established.


\subsection{Runtime Evaluation}
To answer RQ3, runtime is measured across all stages of the XION pipeline.

In the offline phase, the total time required to learn the dependency graph and train the corresponding local predictive models is measured.

For anomaly detection, the inference times of the DP and IF detectors are compared. Inference time is defined as the wall-clock time required to process instances and produce alert decisions.

Finally, the time required to obtain post-alert information is examined. For XION, the time required to compute dependency violation scores over attack instances is measured and reported as the average time per processed instance. For the SHAP baseline, the time
required to compute feature attributions is measured and reported as the average time per explained alert. These measurements reflect different workloads: dependency violation scores are produced as part
of XION's anomaly-detection process, whereas SHAP requires an additional post-hoc attribution step.

\subsection{Sensitivity Analysis}
In addition to the main evaluation, sensitivity
analyses are conducted to examine how the results are affected by selected parameter choices.

For anomaly detection, the dependency violation
threshold $\tau$ and the minimum number of
simultaneously violated features $k$ required to trigger an alert are varied around their default values. This analysis examines how changes in the strictness of the alerting criteria affect detection performance.

For post-alert analysis, the sensitivity of the
temporal stage assignments to the ground-truth-based ES and LS boundaries is examined. Specifically, the ES
boundary $\alpha$ is varied between 2.5\% and
10\%, while the LS boundary $\beta$ is varied between 5\% and 15\%. This analysis assesses whether the observed feature prioritization and temporal patterns remain consistent under alternative definitions of the early and late stages.

\section{Evaluation Outcomes}
\label{sec:results}

This section presents and discusses the results of the learned dependency graph and the experiments addressing the RQs, followed by the sensitivity analysis.

\subsection{Dependency Graph under Benign Conditions}
\label{sec:graph}
Table~\ref{tab:r2_stats} reports the goodness-of-fit of the local feature predictors on benign training and held-out validation data. 
The results show that the learned dependency graphs capture meaningful feature relationships under benign conditions, with high predictive accuracy. The median $R^2$ is 0.99 in both datasets, with similar results on training and validation data, indicating that many learned dependencies generalize to unseen benign instances. More details on the number of neighbors in the learned dependency graphs are provided in
Appendix~\ref{app:graphs}.

\begin{table}[ht]
\small
\centering
\caption{Goodness-of-fit distribution across the local feature predictors on benign training and held-out validation data. Q1 and Q3 denote the first and third quartiles; the maximum $R^2$ was 1.00 in all cases.}
\label{tab:r2_stats}
\begin{tabular}{lccccc}
\toprule
\multicolumn{6}{c}{\textbf{CIC-IDS-2018}} \\
\midrule
Split & Min & Q1 & Median & Q3 & $\%\, R^2 \geq 0.90$ \\
\midrule
Train      & 0.13 & 0.96 & 0.99 & 1.00 & 90.0 \\
Validation & 0.12 & 0.96 & 0.99 & 1.00 & 90.0 \\
\midrule
\multicolumn{6}{c}{\textbf{CIC-UNSW-NB15}} \\
\midrule
Split & Min & Q1 & Median & Q3 & $\%\, R^2 \geq 0.90$ \\
\midrule
Train      & 0.89 & 0.98 & 0.99 & 0.99 & 92.1 \\
Validation & 0.69 & 0.96 & 0.99 & 0.99 & 89.5 \\
\bottomrule
\end{tabular}
\end{table}

To further examine the learned dependency structure, we inspected the strongest learned relationships. These include both closely related or derived features, and relationships across distinct traffic characteristics, such as TCP flags, inter-arrival times, packet lengths, and window sizes. Small subsets of the graphs are shown in Appendix \ref{app:graphs}, as visualizing the complete graphs would be prohibitive.

\begin{table}[H]
\centering
\footnotesize
\caption{Distributional analysis of dependency violations. Median values are reported for benign and attack traffic. All Mann--Whitney U tests yield $p < 0.01$.}
\label{tab:distr}
\begin{tblr}{
  row{1} = {c,font=\bfseries},
  row{3} = {c},
  row{7} = {c,font=\bfseries},
  cell{1}{1} = {c=7}{},
  cell{2}{1} = {r=2}{},
  cell{2}{2} = {c=3}{c},
  cell{2}{5} = {c=3}{c},
  cell{4}{2} = {r=3}{c},
  cell{4}{3} = {c},
  cell{4}{4} = {c},
  cell{4}{5} = {r=3}{c},
  cell{4}{6} = {c},
  cell{4}{7} = {c},
  cell{5}{3} = {c},
  cell{5}{4} = {c},
  cell{5}{6} = {c},
  cell{5}{7} = {c},
  cell{6}{3} = {c},
  cell{6}{4} = {c},
  cell{6}{6} = {c},
  cell{6}{7} = {c},
  cell{7}{1} = {c=7}{},
  cell{8}{1} = {r=2}{},
  cell{8}{2} = {c=3}{c},
  cell{8}{5} = {c=3}{c},
  cell{10}{2} = {r=3}{c},
  cell{10}{3} = {c},
  cell{10}{4} = {c},
  cell{10}{5} = {r=3}{c},
  cell{10}{6} = {c},
  cell{10}{7} = {c},
  cell{11}{3} = {c},
  cell{11}{4} = {c},
  cell{11}{6} = {c},
  cell{11}{7} = {c},
  cell{12}{3} = {c},
  cell{12}{4} = {c},
  cell{12}{6} = {c},
  cell{12}{7} = {c},
  hline{1-2,4,7-8,10,13} = {-}{},
}
CIC-IDS-2018 &          &        &      &          &        &      \\
Class               & $v_{mean}$ &        &      & $n_{viol}$ &        &      \\
                     & Benign   & Attack & KS   & Benign   & Attack & KS   \\
{DDoS \\HTTP}        & 0.29     & 0.46   & 0.70 & 0~       & 2      & 0.71 \\
Slowloris            &          & 1.68   & 0.70 &          & 11     & 0.70 \\
SSH BF               &          & 0.67   & 0.47 &          & 6      & 0.70 \\
CIC-UNSW-NB15    &          &        &      &          &        &      \\
Class              & $v_{mean}$ &        &      & $n_{viol}$ &       &      \\
                     & Benign   & Attack & KS   & Benign~  & Attack & KS   \\
DoS                  & 0.24     & 1.07   & 0.84 & 0        & 8      & 0.84 \\
Exploits             &          & 1.18   & 0.85 &          & 10     & 0.85 \\
Recon                &          & 1.18   & 0.84 &          & 9      & 0.83 
\end{tblr}
\end{table}

\subsection{Dependency Violation as Anomaly Signals}
The results reported below show evidence that dependency violations occur during attacks and provide anomaly signals that are broadly comparable to those of the ML baseline (RQ1).

\subsubsection{Dependency Violation Distribution Analysis}
\label{sub:distr}
Table~\ref{tab:distr} summarizes the distribution of dependency violations across the evaluated attack scenarios. 
The mean violation score ($v_{mean}$) columns show larger median violation scores for attack flows, while the $n_{viol}$ columns indicate that more feature relationships are simultaneously violated during attacks. Higher values of either metric for attack flows indicate stronger deviations from normal instances. 
High KS values for $n_{viol}$ suggest that attacks
frequently produce simultaneous violations across
multiple feature dependencies. For all attacks, the Mann--Whitney U tests yield statistically significant differences ($p < 0.01$), indicating that the observed shifts in dependency violation behavior are unlikely to arise from random variation. 

\subsubsection{Anomaly Detection Performance}
\label{sub:ad}

Table \ref{tab:detector_comparison} reports standard detection metrics in the $[0,1]$ range for the DP detector and the IF baseline.  

\begin{table}[H]
\small
\centering
\caption{Detection performance per attack for the DP detector (DP) and Isolation Forest (IF) across datasets.}
\label{tab:detector_comparison}
\begin{tblr}{
  row{1} = {c,font=\bfseries},
  row{2} = {font=\bfseries},
  row{9} = {c,font=\bfseries},
  row{10} = {font=\bfseries},
  cell{1}{1} = {c=7}{},
  cell{3}{1} = {r=3}{},
  cell{3}{6} = {r=3}{},
  cell{3}{7} = {r=3}{},
  cell{6}{1} = {r=3}{},
  cell{6}{6} = {r=3}{},
  cell{6}{7} = {r=3}{},
  cell{9}{1} = {c=7}{},
  cell{11}{1} = {r=3}{},
  cell{11}{6} = {r=3}{},
  cell{11}{7} = {r=3}{},
  cell{14}{1} = {r=3}{},
  cell{14}{6} = {r=3}{},
  cell{14}{7} = {r=3}{},
  hline{1-3,6,9-11,14,17} = {-}{},
}
CIC-IDS-2018 &                &      &      &      &       &       \\
Detector             & Class         & Rec. & BA   & MCC  & FPR~  & Spec. \\
DP                   & {DDoS \\ HTTP} & 0.94 & 0.86 & 0.73 & 0.22~ & 0.78  \\
                     & Slowloris      & 0.80 & 0.79 & 0.29 &       &       \\
                     & SSH BF         & 0.50 & 0.64 & 0.30 &       &       \\
IF                   & {DDoS \\ HTTP} & 0.84 & 0.85 & 0.64 & 0.14~ & 0.86  \\
                     & Slowloris      & 0.78 & 0.82 & 0.38 &       &       \\
                     & SSH BF         & 0.51 & 0.69 & 0.40 &       &       \\
CIC-UNSW-NB15    &                &      &      &      &       &       \\
Detector             & Class         & Rec. & BA   & MCC  & FPR   & Spec. \\
DP                   & DoS            & 0.99 & 0.90 & 0.30 & 0.17  & 0.82  \\
                     & Exploits       & 0.99 & 0.90 & 0.61 &       &       \\
                     & Recon          & 1.00 & 0.91 & 0.51 &       &       \\
IF                   & DoS            & 0.71 & 0.80 & 0.27 & 0.10~ & 0.89  \\
                     & Exploits       & 0.72 & 0.81 & 0.54 &       &       \\
                     & Recon          & 0.39 & 0.64 & 0.23  &       &       
\end{tblr}
\end{table}

Overall, the DP detector achieves high recall across most attack categories, while IF performance is more variable, particularly for Reconnaissance traffic in CIC-UNSW-NB15. 

The ROC and precision--recall analyses further show broadly comparable discrimination between the two detectors on CIC-IDS-2018, although IF achieves higher AUPRC for Slowloris. On CIC-UNSW-NB15, the DP detector consistently achieves higher AUROC and AUPRC than IF. Full curves and metric values are reported in Appendix~\ref{app:roc_pr}.

Notably, the DP detector achieves balanced accuracy comparable to IF across most attack scenarios, although generally with higher FP rates. To analyze this behavior, we have inspected the FPs and compared them with the correctly flagged benign samples, revealing deviations. In CIC-IDS-2018, these deviations are related to timing and activity characteristics, whereas in CIC-UNSW-NB15 they more prominently involve packet and segment sizes, together with flag-related features. This is expected, since the DP detector explicitly reacts to violations of individual feature relationships and may therefore be more sensitive to benign fluctuations that locally deviate from the learned dependency structure.
\begin{figure}[H]
    \centering

    \begin{minipage}[t]{0.48\textwidth}
        \centering
        \textbf{CIC-IDS-2018}

        \vspace{0.3em}

        \begin{subfigure}[b]{0.95\textwidth}
            \centering
            \includegraphics[width=\textwidth]{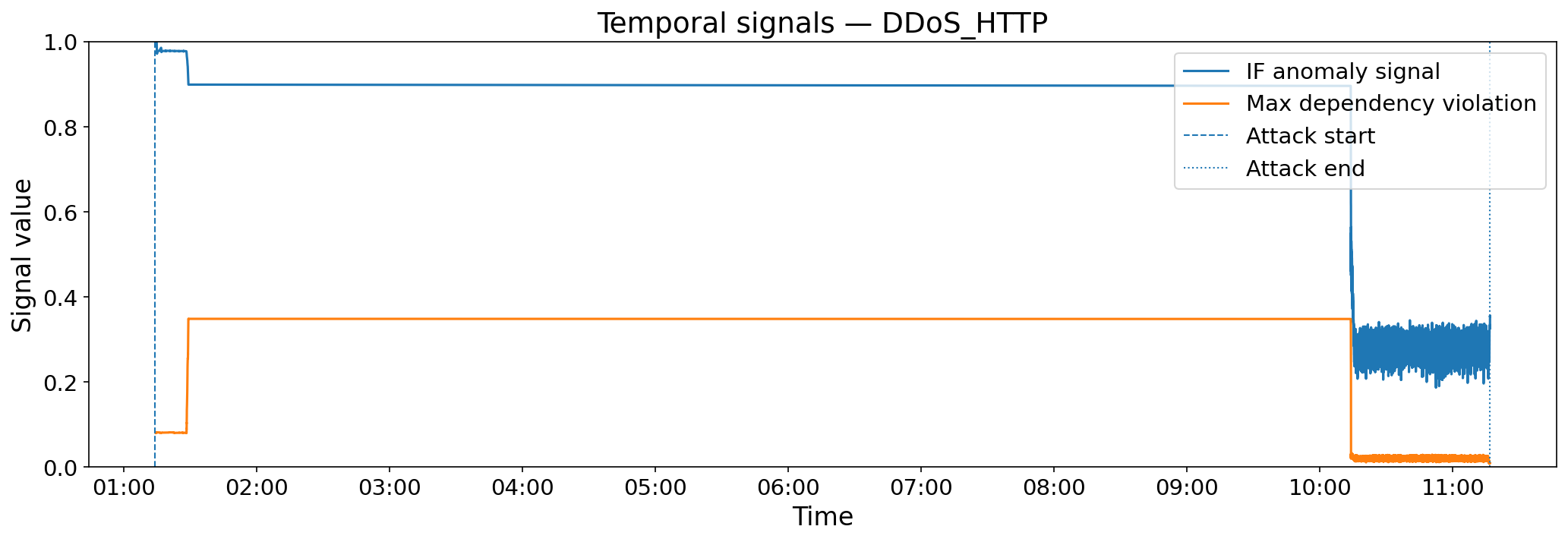}
            \caption{DDoS HTTP}
            \label{fig:ddos}
        \end{subfigure}

        \vspace{0.3em}

        \begin{subfigure}[b]{0.95\textwidth}
            \centering
            \includegraphics[width=\textwidth]{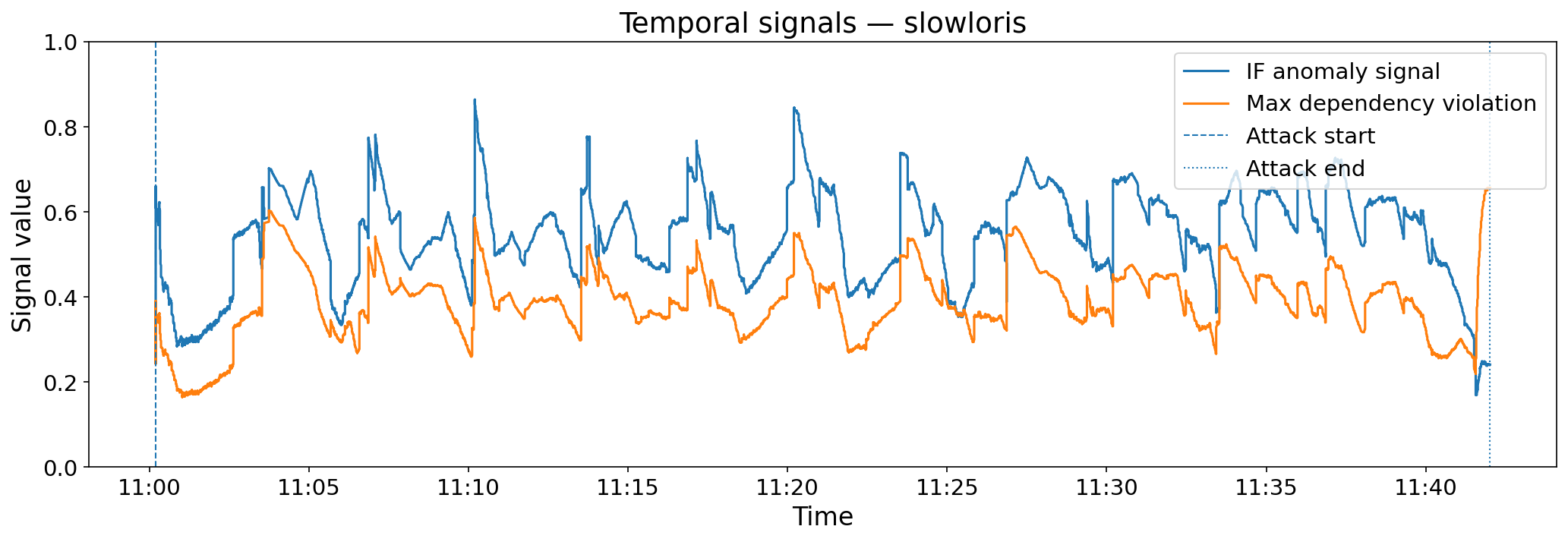}
            \caption{Slowloris}
            \label{fig:slowloris}
        \end{subfigure}

        \vspace{0.3em}

        \begin{subfigure}[b]{0.95\textwidth}
            \centering
            \includegraphics[width=\textwidth]{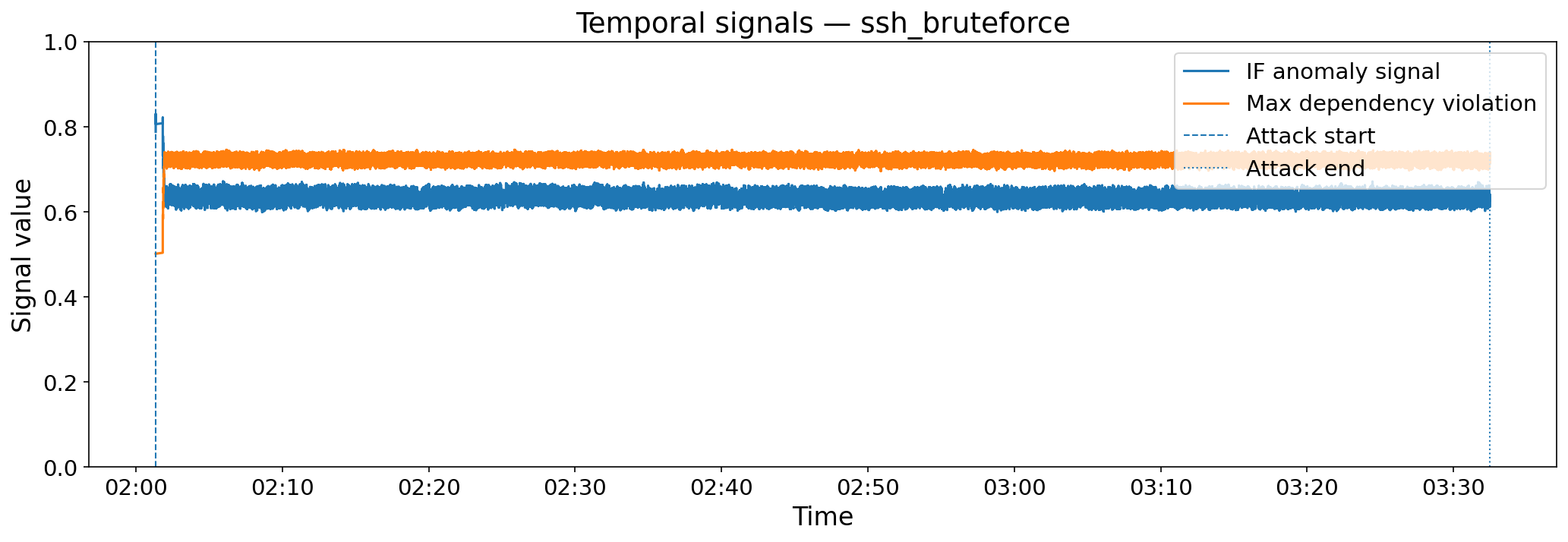}
            \caption{SSH brute-force}
            \label{fig:ssh_bruteforce}
        \end{subfigure}
    \end{minipage}
    \hfill
    \begin{minipage}[t]{0.48\textwidth}
        \centering
        \textbf{CIC-UNSW-NB15}

        \vspace{0.3em}

        \begin{subfigure}[b]{0.95\textwidth}
            \centering
            \includegraphics[width=\textwidth]{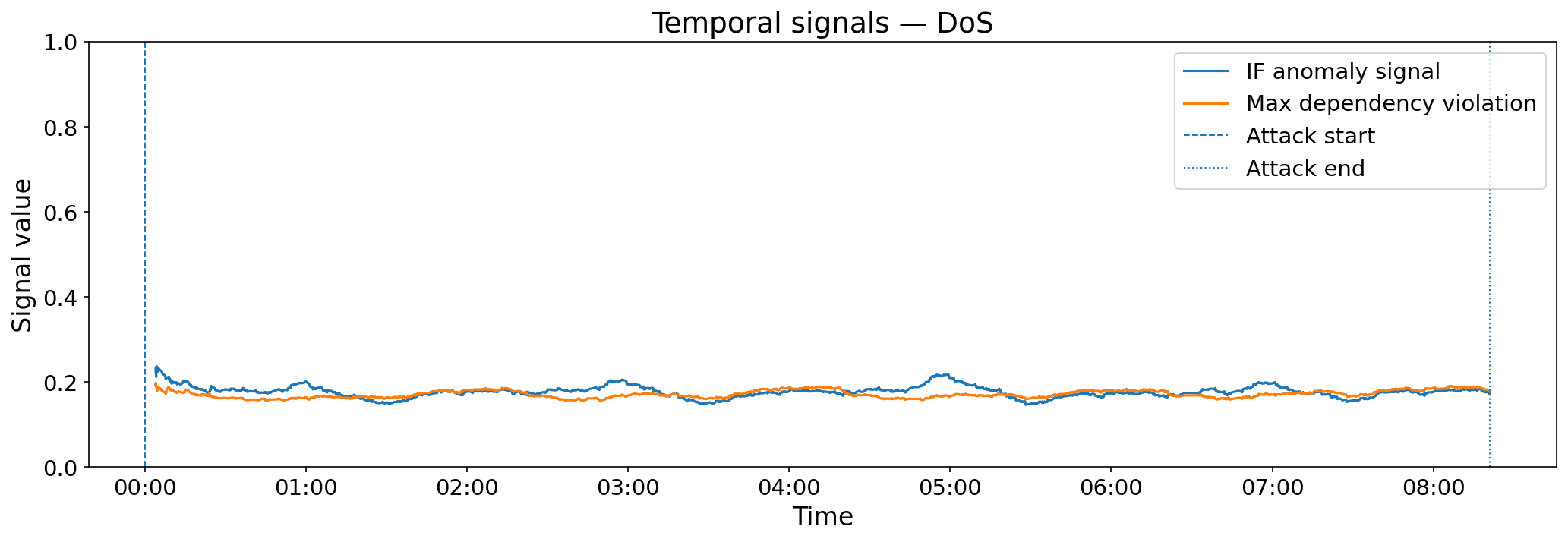}
            \caption{DoS}
            \label{fig:dos}
        \end{subfigure}

        \vspace{0.3em}

        \begin{subfigure}[b]{0.95\textwidth}
            \centering
            \includegraphics[width=\textwidth]{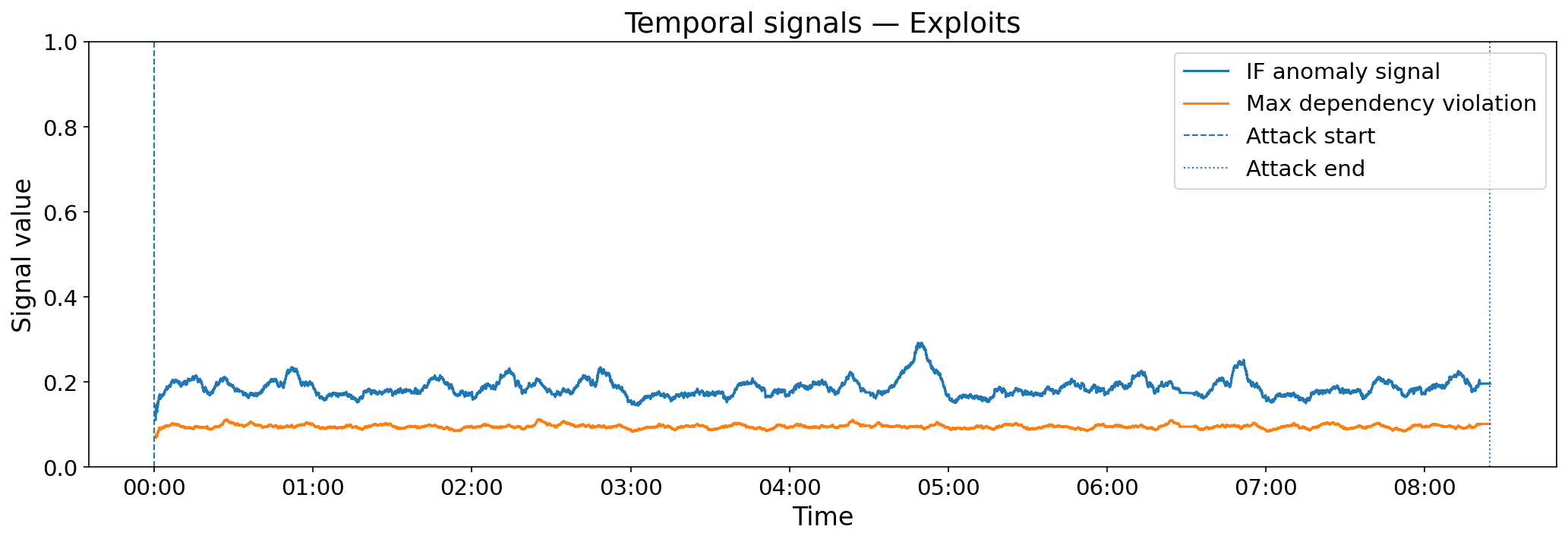}
            \caption{Exploits}
            \label{fig:exploits}
        \end{subfigure}

        \vspace{0.3em}

        \begin{subfigure}[b]{0.95\textwidth}
            \centering
            \includegraphics[width=\textwidth]{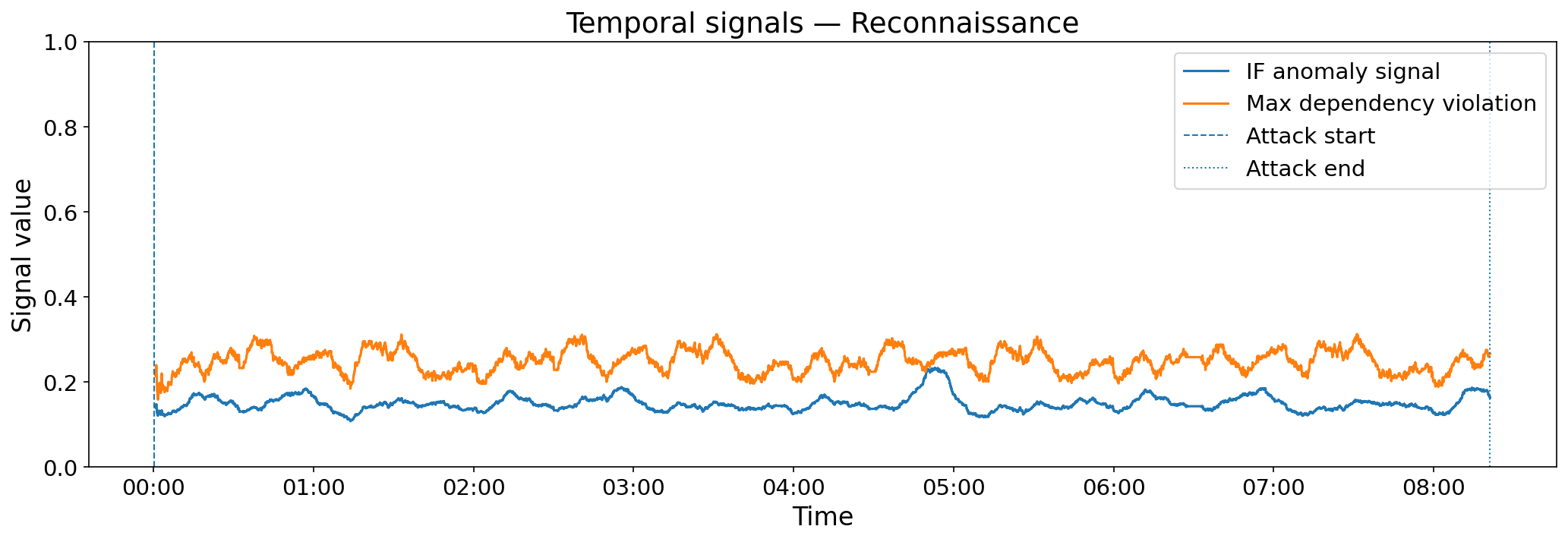}
            \caption{Reconnaissance}
            \label{fig:recon}
        \end{subfigure}
    \end{minipage}

    \caption{
    Temporal comparison of the IF anomaly score (blue) and dependency
    violation signal (orange) across all attack scenarios. The attack
    timeline is expressed as time of day. To enable temporal comparison,
    both signals are normalized to the $[0,1]$ range.
    }
    \label{fig:overlap_all}
\end{figure}
\subsubsection{Temporal Overlap}
Figure \ref{fig:overlap_all} compares the normalized IF anomaly scores (blue) and maximum dependency violation scores (orange) over time. 
The dashed and dotted vertical lines indicate attack onset and end, respectively, based on the first and last instances labeled as malicious in the datasets'ground-truth annotations.

The signals exhibit similar temporal patterns across most attacks, although some differences
emerge. The signals closely follow each other for DoS. During DDoS HTTP, the dependency violation signal shows a similar, relatively stable signal compared to IF. For Slowloris and SSH brute force, both signals remain fairly steady, with minor fluctuations. 
For Exploits and Reconnaissance, dependency violations are more persistent despite comparatively weaker or more variable IF scores.

The DP detector and IF produce highly overlapping anomaly signals despite relying on different modeling assumptions. Inspection of the alert timestamps indicates that both detectors react at the first instance labeled as malicious,
at the temporal granularity of the datasets.

\begin{table}
\small
\centering
\caption{Qualitative evaluation of post-alert evidence against the literature-grounded behavioral reference reported in Appendix~\ref {app:literature}.}
\label{tab:pa_evaluation}
\footnotesize
\begin{tblr}{
  row{1} = {font=\bfseries},
  hline{1-2,5,8} = {-}{},
}
Attack         & XION evidence and assessment                                                                                                                                                                                  & SHAP evidence and assessment                                                                                                                                                                                                          \\
{DDoS \\ HTTP} & {\textbf{Supported.} Broad violations\\involving volume, packet size, \\IAT, and TCP characteristics \\emerge predominantly \\in ES, consistent with the \\widespread traffic disruption of \\HTTP flooding.} & {\textbf{Partially supported.}~Attribution\\in ES emphasizes forward traffic \\volume,while later attribution shifts \\toward TCP, packet-size, and IAT \\features, providing partial\\evidence of the defining flooding \\behavior.} \\
Slowloris      & {\textbf{Supported.} Timing, activity, and\\packet-size relationships deviate \\progressively across the attack, \\consistent with slow, persistent \\connection activity.}                                   & {\textbf{Supported.} IAT and idle-time\\characteristics are prioritized in \\ES and MS and remain present \\in LS, consistent with the \\behavior of Slowloris.}                                                                      \\
SSH BF         & {\textbf{Partially supported.} Connection\\-level violations emerge only in \\ES, indicating localized disruption, \\but do not directly reflect \\authentication attempts.}                                  & {\textbf{Partially supported.} Connection\\-level characteristics are prioritized \\throughout the attack but do not \\directly reflect authentication \\attempts.}                                                                   \\
DoS            & {\textbf{Supported.} Broad traffic, packet,\\timing, and TCP relationships \\predominantly deviate in ES, \\consistent with substantial\\traffic disruption during DoS.}                                      & {\textbf{Partially supported.} Packet/segment,\\TCP, timing, and rate features are \\highlighted across all stages, with little \\temporal differentiation and partial \\evidence of the attack behavior.}                            \\
Exploits       & {\textbf{Partially supported.} Broad\\packet, traffic, timing, and TCP \\relationships deviate in ES, but \\cannot be associated with a \\specific exploitation mechanism.}                                   & {\textbf{Not supported.} Generic packet, TCP,\\and timing characteristics persist across \\all stages, without evidence of a specific \\exploitation mechanism.}                                                                      \\
Recon          & {\textbf{Partially supported.} Throughput,\\protocol, packet, TCP, and timing \\relationships predominantly deviate \\in ES, but do not identify \\a specific reconnaissance\\mechanism.}                     & {\textbf{Partially supported.} Rate, protocol,\\segment-size, and timing features are \\prioritized throughout the attack, \\consistent with probing traffic but not a \\specific reconnaissance mechanism.}                          
\end{tblr}
\end{table}

\subsection{Dependency Violation for Post-Alert Analysis}
\label{sub:pa}
Table~\ref{tab:pa_evaluation} reports the qualitative evaluation of XION and the SHAP baseline against the literature-grounded behavioral reference (see Appendix \ref{app:literature}). For the six classes of attacks, overall, there is agreement between the two analysis methods (XION and SHAP) and the ground truth based on the literature. However, for three of the attack classes, XION provides better support compared to the partial support provided by SHAP. In one case, SHAP does not provide any support, as the highlighted features are not aligned with the typical behaviour of a flow under the attack (Exploits).

For DDoS HTTP and DoS, XION identifies widespread disruptions in traffic volume, packet, timing, and TCP dependencies, predominantly beginning in an early stage (ES).
SHAP highlights relevant features but provides less temporal differentiation across attack stages.
For Slowloris, both approaches align with the behavioral reference. XION captures timing-and activity-related dependencies first deviating at different stages, while SHAP consistently prioritizes IAT and idle-time features.

The evidence is less specific for SSH brute force, Exploits, and Reconnaissance. This is partly expected from the flow-level data: the defining behavior of SSH brute force concerns authentication attempts, while Exploits and Reconnaissance represent broad attack categories that do not imply a single flow-level signature. However, even when the affected features cannot be mapped to a specific attack mechanism, their dependency and temporal context provide additional information for post-alert investigation.


Across both datasets, the neighbor-ordering analysis (Algorithm~\ref{alg:pa}) helps prioritize features by identifying those whose dependencies are violated earlier than those of their neighbors. For example, during DDoS HTTP and DoS, forward-packet counts deviate before neighboring timing and packet-length features. These earlier deviations provide a starting point for investigating how the anomaly manifests across related traffic features.

The qualitative evaluation shows that dependency violations help prioritize features aligned with the expected behavior of different attack categories, better than what SHAP does (RQ2), while providing additional context for post-alert investigation that is not directly available from aggregate anomaly scores nor post-hoc feature attribution. 


\subsection{Runtime Evaluation}
\label{sub:time}

The results indicate that dependency-based detection and post-alert analysis can be performed with low execution times, with dependency violation scores computed during detection rather than through an additional post-hoc attribution step (RQ3).

In the offline phase, the total time required to learn the dependency graph and fit all local predictive models was 48.85 seconds for CIC-IDS-2018 and 76.51 seconds for CIC-UNSW-NB15. Notably, although CIC-UNSW-NB15 contains twice as many benign samples, the runtime did not increase proportionally.

\begin{figure}[ht]
    \centering

    \begin{subfigure}[b]{0.48\textwidth}
        \centering
        \includegraphics[width=\textwidth]{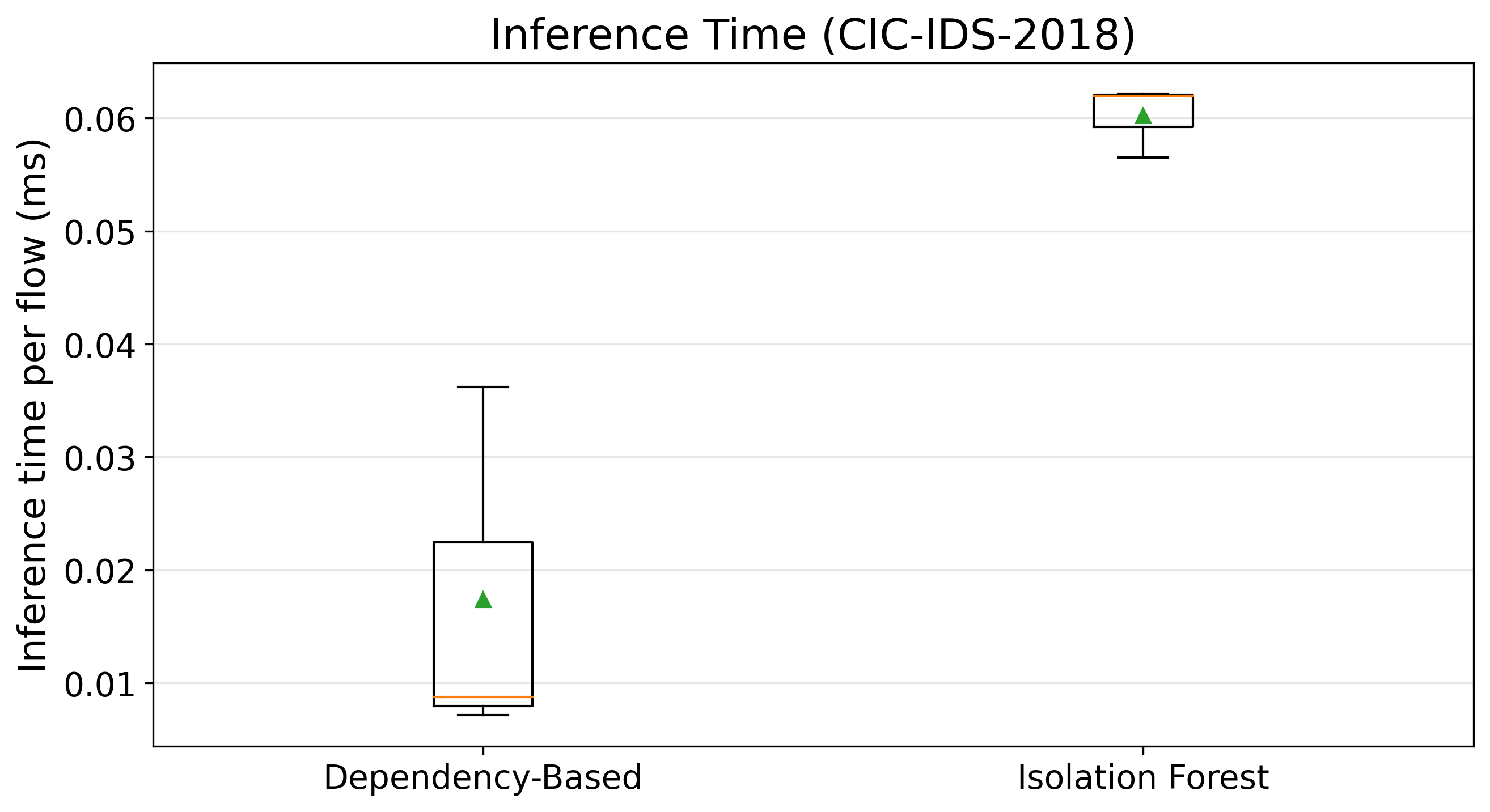}
        \caption{CIC-IDS-2018}
        \label{fig:runtime_cicids}
    \end{subfigure}
    \begin{subfigure}[b]{0.48\textwidth}
        \centering
        \includegraphics[width=\textwidth]{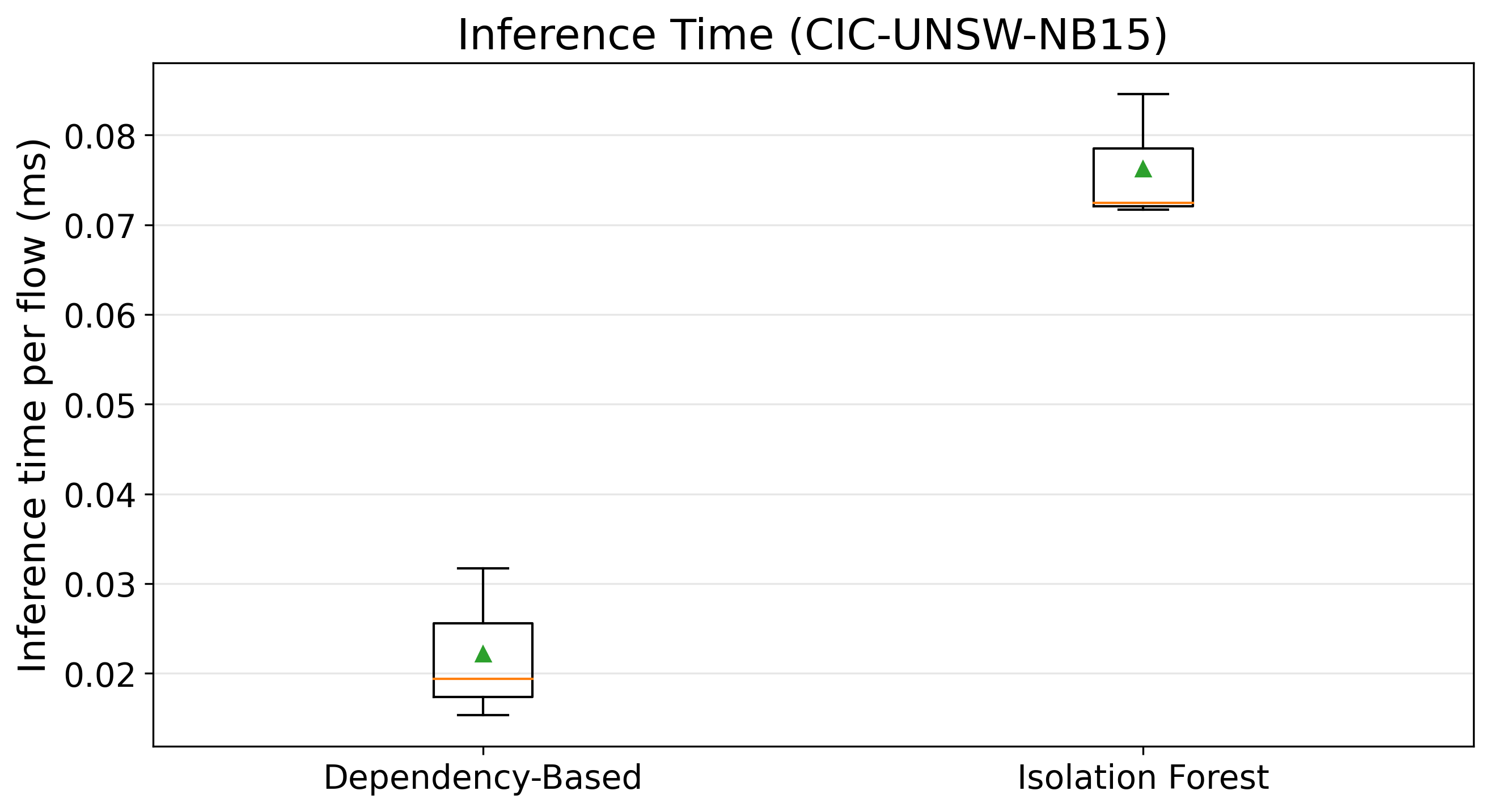}
        \caption{CIC-UNSW-NB15}
        \label{fig:runtime_unsw}
    \end{subfigure}

    \caption{
    Distribution of average per-flow inference time across attack datasets for the DP and IF detectors.}
    \label{fig:runtime}
\end{figure}

Figure \ref{fig:runtime} reports the distribution of average per-flow inference time across all attack datasets for the DP and IF detectors. These measurements correspond to detector processing time per network flow; the total attack duration varies from minutes to hours depending on the scenario. In each boxplot, the central line denotes the median, the box spans the interquartile range, the whiskers represent the non-outlier range, and the green triangle indicates the mean inference time.

Across the evaluated scenarios, violation computation for XION required approximately 0.008--0.02~ms per processed flow. In contrast, SHAP constitutes an additional post-hoc computation applied to IF alerts, requiring approximately 0.1~s per explained alert.

\subsection{Sensitivity Analysis}
The sensitivity analysis reveals a trade-off between recall and false-positive rate when varying $\tau$ and $k$. Detection performance remains relatively stable across the tested configurations on CIC-UNSW-NB15, while greater sensitivity is observed
on CIC-IDS-2018, particularly for SSH brute force (Appendix~\ref{app:sensitivity}).

The temporal classification of dependency violations is largely robust to changes in the ES and LS boundaries. First-violation stage assignments remain unchanged across all tested boundaries on CIC-UNSW-NB15. On CIC-IDS-2018, some features shift
between adjacent stages, particularly from MS to LS for Slowloris, without substantially altering the overall temporal patterns.

\section{Limitations and future work}
\label{sec:lim_fw}
Despite the promising results, some limitations should be considered when interpreting the findings.
Our evaluation relies on the pre-collected datasets, which may not fully capture the diversity and variability of real-world network traffic.

Recent research shows concern relative to labeling standard practice
\cite{braun2024understanding,botacin2025towards}. Labels may contain inaccuracies or ambiguities, and some samples labeled as malicious may resemble benign behavior, or vice versa, potentially affecting training and evaluation outcomes. 
Moreover, previous studies have identified issues in traffic capture, flow construction, and labeling in CIC datasets, some associated with CICFlowMeter\footnote{\url{https://github.com/ahlashkari/CICFlowMeter}} \cite{lanvin2022errors,engelen2021troubleshooting}. Such artifacts may impact feature values and the dependencies learned between them. 
However, these limitations are common in IDS literature, where such datasets remain widely used as de facto benchmarks.

XION is limited to numerical features and operates on aggregated features extracted by CICFlowMeter. While this representation is standard in network intrusion detection, it does not capture packet-level or protocol-specific information that may provide additional evidence for detecting or interpreting certain attacks. At the same time, XION does not rely on packet payload contents, suggesting potential applicability to encrypted traffic where payload visibility is limited. The work would benefit from being applied and tested on encrypted traffic.

Several directions can extend the work and further validate XION. To address the shortcomings of publicly available datasets, recent literature has proposed approaches to generate synthetic
attack traffic \cite{10628730}, which could provide a first step. A stronger validation, however, would involve a controlled testbed where the causes of anomalies are explicitly known. 

Future work will investigate the application of dependency-based analysis to multi-step attack scenarios. In complex attack chains, individual stages may not produce strong anomaly signals when considered in isolation. Our ongoing work studies dependency violations as a complement to structured representations of attack progression (e.g., AI planning models), to capture how anomalies evolve across stages and identify transition points in the attack timeline.

Another important direction is evaluating the robustness of the dependency model under concept drift, where the benign data distribution changes over time. Understanding how such shifts affect learned feature relationships would be valuable for assessing the adaptability of XION in real-world environments.

\section{Conclusion}
\label{sec:conclusion}

In this work, we explored whether anomalies in network traffic can be understood as disruptions of the normal relationships between flow-level features. By learning conditional dependencies from benign traffic only, we modeled the expected structure of interactions among network-flow characteristics and identified attacks through violations of these relationships.

Experiments on CIC-IDS-2018 and CIC-UNSW-NB15 datasets show that dependency violations emerge consistently during attack scenarios and generate anomaly signals that are comparable with those of a baseline IF detector. 

Across both datasets, XION captures distinct temporal and structural disruption patterns associated with different attack behaviors. 

Beyond detection, XION provides additional interpretability by revealing \emph{which} relationships break, \emph{when} they break, and \emph{how} violations evolve over time.

These findings suggest that dependency violations capture meaningful structural changes in network behavior and can assist post-alert investigation by prioritizing potentially relevant features and relationships. 
More broadly, modern network monitoring relies on multiple indicators and analytical tools to detect and investigate attacks: 
dependency modeling can play a complementary role within this toolbox by providing structural insight into anomalous traffic behavior without requiring explicit causal models or domain-specific rules.

\section*{Acknowledgment}
This work was carried out within the NEST project $AIR^2$, which is partially supported by the Wallenberg AI, Autonomous Systems and Software Program (WASP) funded by the Knut and Alice Wallenberg Foundation. 

\bibliographystyle{elsarticle-num}
\bibliography{reference.bib}

\appendix

\section{Datasets Details}
\label{app:dataset}

Attacks descriptions:
\begin{itemize}
    \item \textbf{HTTP flood Distributed Denial of Service (DDoS)}: an attack in which multiple compromised machines overwhelm a target web server with a massive volume of HTTP requests, exhausting network or server resources and preventing legitimate users from accessing the service.
    \item \textbf{Slowloris}: a stealthier low-rate DoS attack that keeps many HTTP connections open by sending incomplete requests slowly, forcing the server to maintain half-open sessions until its connection pool is exhausted.
    \item \textbf{SSH brute-force attack}: an attack that repeatedly attempts different username and password combinations against the SSH service in order to gain unauthorized access through credential guessing. 
    \item \textbf{DoS}: an attack aimed at disrupting the availability of a system or service by overwhelming it with excessive traffic or resource-intensive requests.
    \item \textbf{Exploit}: an attack that takes advantage of software vulnerabilities or security flaws in applications, operating systems, or network services in order to execute unauthorized actions or gain access to a system.
    \item \textbf{Reconnaissance}: an attack focused on gathering information about a target system or network, such as open ports, running services, or vulnerabilities, to prepare for future intrusions. 
\end{itemize}

For modeling, starting from all 84 raw CICFlowMeter columns, five non-feature fields (Flow ID, Source IP, Destination IP, Timestamp, and Label) were excluded, resulting in a 79-dimensional feature space, all numerical. Attack and benign files were stored with 81 columns in total, comprising Timestamp, the 79 feature columns, and Label. 

All attack traces were organized chronologically based on timestamp order. In CIC-IDS-2018, each attack is collected in a separate calendar day. In CIC-UNSW-NB15, attacks span multiple days. Hence, the calendar day containing the longest duration of malicious activity is selected for observing each attack.

Table \ref{tab:dataset_summary} summarizes the dimensionality and purpose of each subset across training, validation, and test splits.

\begin{table}[ht]
\centering
\small
\caption{Datasets composition and dimensionality.}
\label{tab:dataset_summary}
\begin{tblr}{
  row{1} = {font=\bfseries},
  column{3} = {r},
  cell{2}{1} = {r=6}{},
  cell{8}{1} = {r=6}{},
  hline{1-2,8,14} = {-}{},
}
Dataset               & Class             & Samples   \\
CIC-IDS-2018            & Train Benign      & 1,000,000 \\
                      & Validation Benign & 200,000   \\
                      & Test Benign       & 200,000   \\
                      & DDoS HTTP         & 576,191   \\
                      & SSH Brute-force    & 187,589   \\
                      & Slowloris         & 10,990    \\
CIC-UNSW-NB15      & Train Benign      & 2,000,000 \\
                      & Validation Benign & 200,000   \\
                      & Test Benign       & 200,000   \\
                      & DoS               & 4467      \\
                      & Exploits          & 30,951    \\
                      & Reconnaissance    & 16,735    
\end{tblr}
\end{table}

\newpage

\section{Learned Dependency Graphs}
\label{app:graphs}
\begin{figure}[ht]
    \centering

    \begin{subfigure}[b]{0.68\textwidth}
        \centering
        \includegraphics[width=\textwidth]{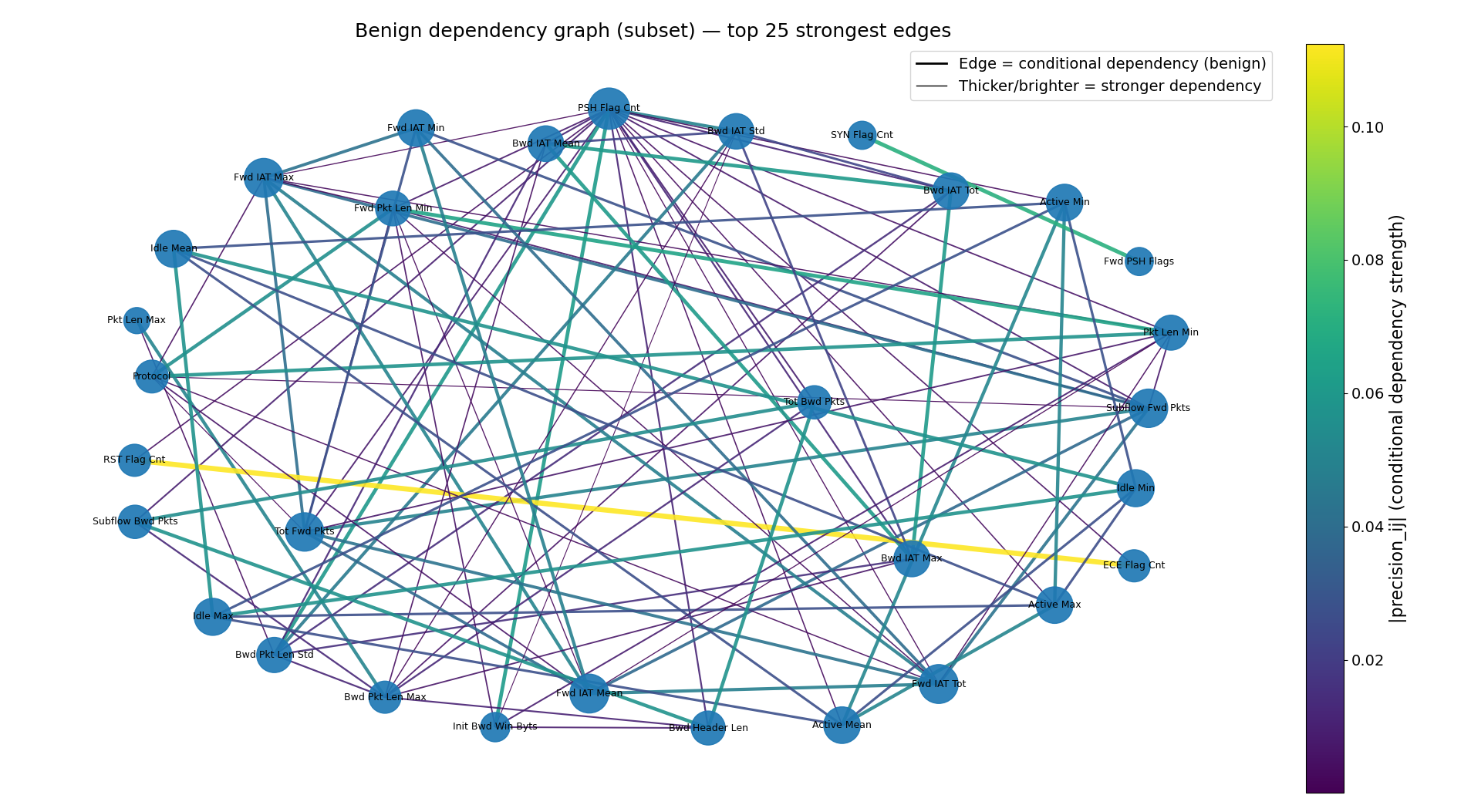}
        \caption{CIC-IDS-2018}
    \end{subfigure}
    \begin{subfigure}[b]{0.68\textwidth}
        \centering
        \includegraphics[width=\textwidth]{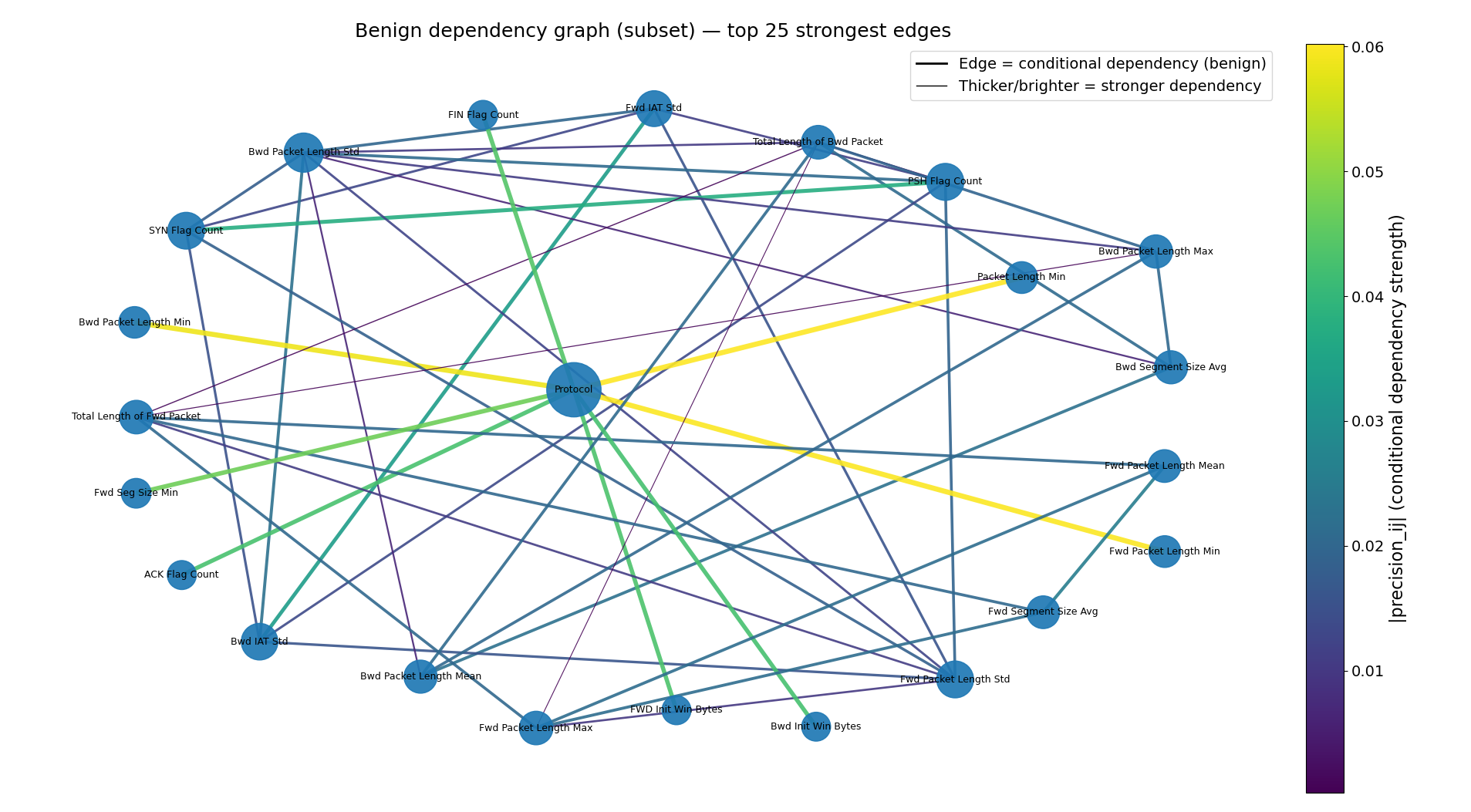}
        \caption{CIC-UNSW-NB15}
    \end{subfigure}

    \caption{
    Example subgraphs extracted from the learned dependency structures.
    }
    \label{fig:model}
\end{figure}

After the pre-processing, both datasets contained 79 features. For CIC-IDS-2018, 69 had neighbors; the number of neighbors in the learned dependency graph ranged from 0 to 29, with a median of 11 neighbors per feature. In CIC-UNSW-NB15, the graph included 70 features with neighbors. Each feature had 0 to 26 neighbors, with a median of 1.
Figure \ref{fig:model} shows two small subsets of the graphs for CIC-IDS-2018 and CIC-UNSW-NB15, respectively, with only the strongest relationships of both graphs included for the sake of intelligibility. 

\newpage

\section{ROC and Precision--Recall Analysis}
\label{app:roc_pr}

\begin{figure*}[ht]
    \centering
    \begin{subfigure}{\textwidth}
        \centering
        \includegraphics[width=\linewidth]{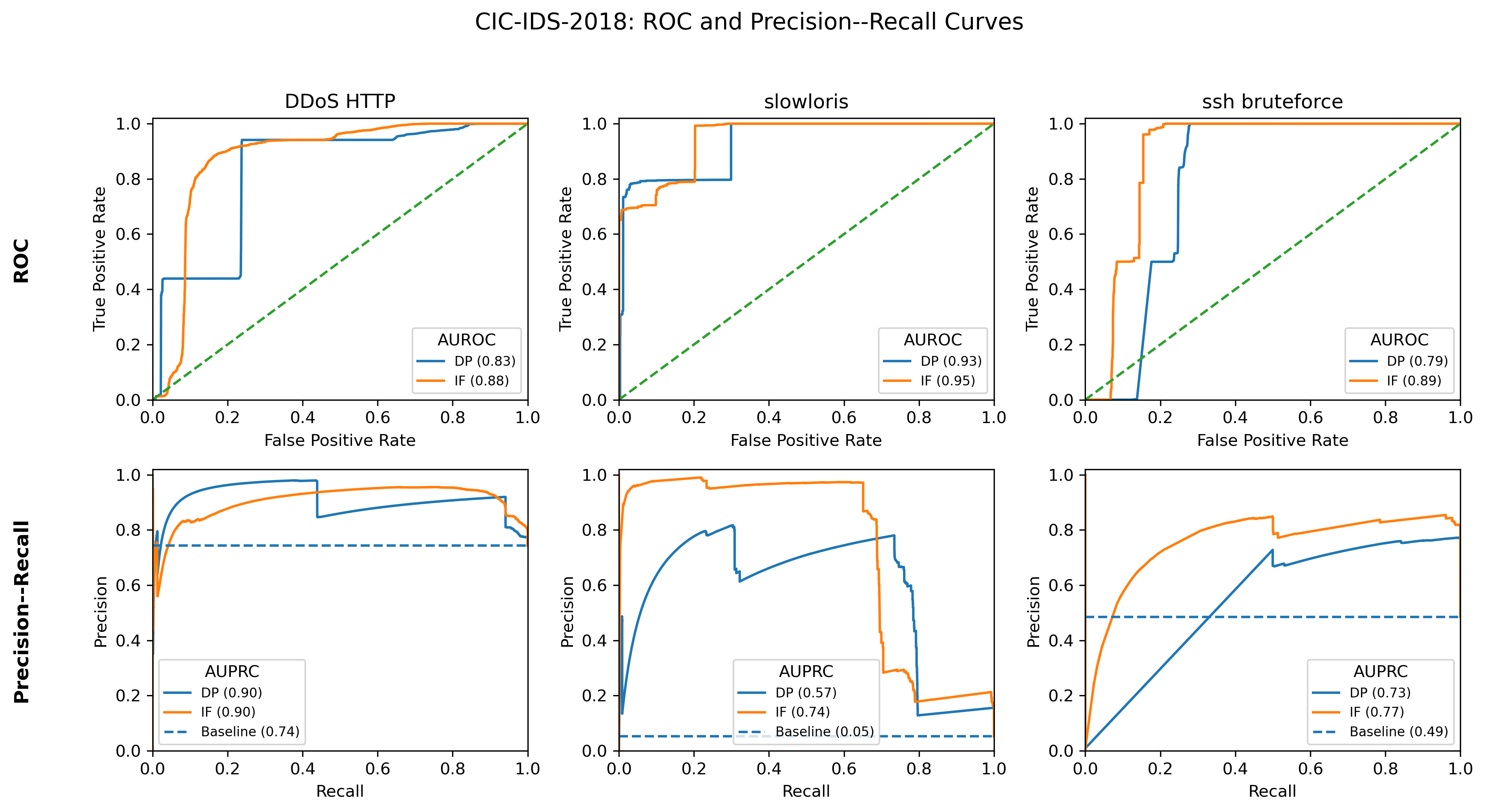}
        \caption{CIC-IDS-2018}
        \label{fig:roc_pr_cicids}
    \end{subfigure}

    \vspace{0.5em}

    \begin{subfigure}{\textwidth}
        \centering
        \includegraphics[width=\linewidth]{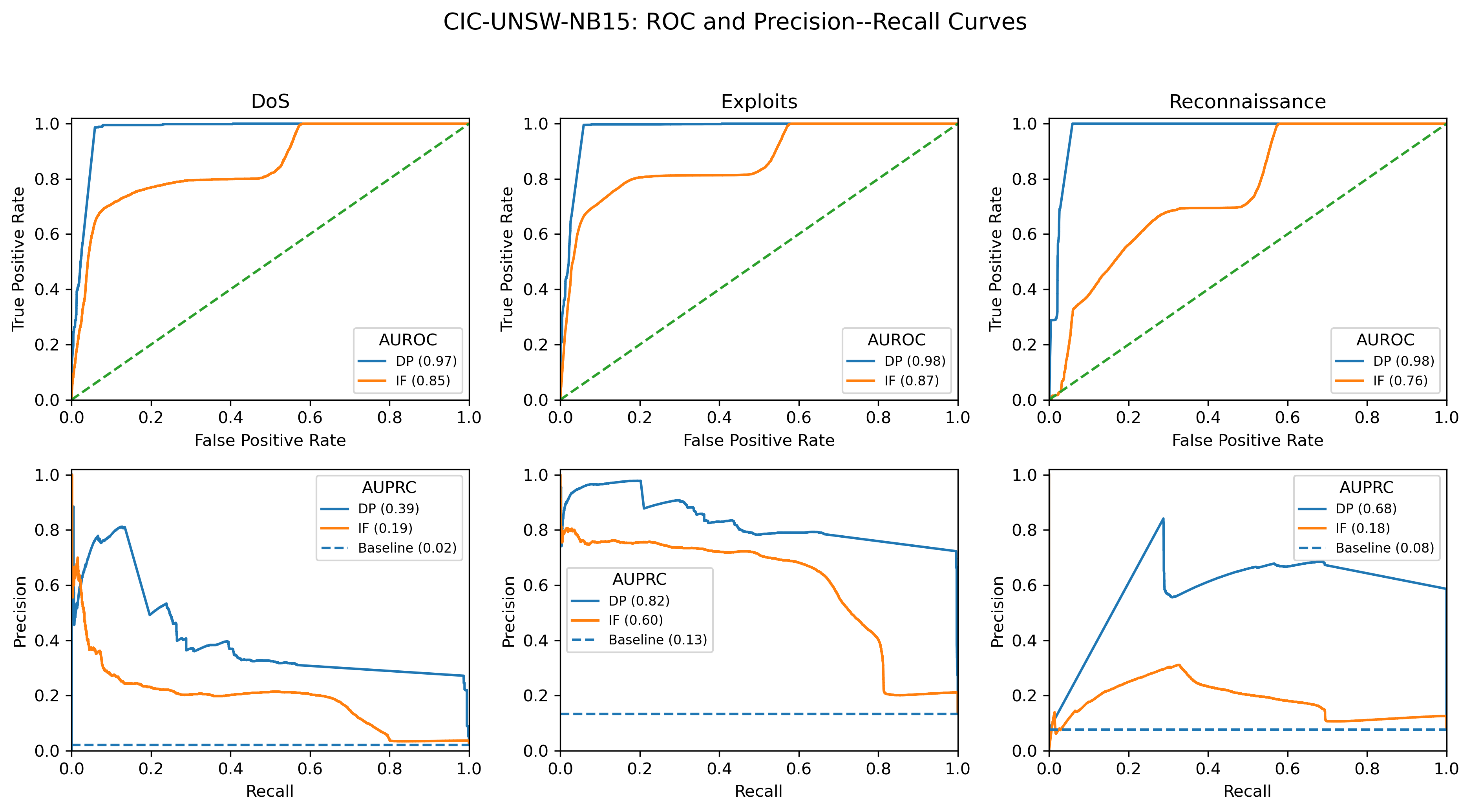}
        \caption{CIC-UNSW-NB15}
        \label{fig:roc_pr_unsw}
    \end{subfigure}

    \caption{ROC and precision--recall curves for DP and IF on
    (a) CIC-IDS-2018 and (b) CIC-UNSW-NB15.
    The dashed horizontal lines in the precision--recall plots
    represent the no-skill baseline, corresponding to the
    prevalence of the respective attack class.}
    \label{fig:roc_pr}
\end{figure*}

Table \ref{tab:roc_pr} reports AUROC and AUPRC for both detectors across all attack scenarios. Figures \ref{fig:roc_pr_cicids} and \ref{fig:roc_pr_unsw} show the corresponding ROC and precision--recall curves. 
Higher AUROC and AUPRC values indicate better discrimination across decision thresholds. An AUROC of 0.5 corresponds to random ranking, while the baseline AUPRC depends on the proportion of attack samples in the evaluated benign--attack pair.

\begin{table}[ht]
\centering
\small
\caption{Threshold-independent detection performance of DP and IF.}
\label{tab:roc_pr}

\setlength{\tabcolsep}{4pt}
\begin{tabular}{lcccc}
\toprule
\multicolumn{5}{c}{\textbf{CIC-IDS-2018}} \\
\midrule
& \multicolumn{2}{c}{AUROC} & \multicolumn{2}{c}{AUPRC} \\
\cmidrule(lr){2-3}\cmidrule(lr){4-5}
Attack & DP & IF & DP & IF \\
\midrule
DDoS HTTP & 0.83 & 0.88 & 0.90 & 0.90 \\
Slowloris & 0.93 & 0.95 & 0.57 & 0.74 \\
SSH BF    & 0.79 & 0.89 & 0.73 & 0.77 \\
\midrule
\multicolumn{5}{c}{\textbf{CIC-UNSW-NB15}} \\
\midrule
& \multicolumn{2}{c}{AUROC} & \multicolumn{2}{c}{AUPRC} \\
\cmidrule(lr){2-3}\cmidrule(lr){4-5}
Attack & DP & IF & DP & IF \\
\midrule
DoS      & 0.97 & 0.85 & 0.39 & 0.19 \\
Exploits & 0.98 & 0.87 & 0.82 & 0.60 \\
Recon.   & 0.98 & 0.76 & 0.68 & 0.18 \\
\bottomrule
\end{tabular}
\end{table}

\section{Literature-base for Post-Alert Analysis}
\label{app:literature}
Table~\ref{tab:pa_grounding} summarizes the literature-grounded
behavioral reference used for the qualitative evaluation of the
post-alert evidence.

\begin{table}[ht]
\centering
\caption{Literature- and dataset-grounded behavioral reference used for post-alert evaluation.}
\label{tab:pa_grounding}
\begin{tblr}{
  row{1} = {font=\bfseries},
  cell{2}{1} = {r=3}{},
  cell{5}{1} = {r=3}{},
  hline{1-2,5,8} = {-}{},
}
Dataset      & Attack        & Literature-grounded attack behavior                                                                                                                                                                                                                                                                                                                                                                                     \\
CIC-IDS-2018 & {DDoS \\HTTP} & {A distributed HTTP flooding attack where\\multiple attacking machines direct HTTP \\requests toward the victim. DDoS flooding \\is characterized by coordinated, large-scale \\activity intended to disrupt legitimate access \\to the targeted service \cite{Sharafaldin2018TowardGA,zargar2013survey}.}                                                                                                                           \\
             & Slowloris     & {A low-rate HTTP denial-of-service attack that \\establishes TCP connections and keeps them \\open by periodically sending valid but incomplete \\HTTP requests. The attack relies on persistent\\connections and temporally dispersed activity \\rather than conventiona lhigh-volume flooding\cite{Sharafaldin2018TowardGA,zhijun2020low}.}                                                                                        \\
             & SSH BF        & {A dictionary-based brute-force attack that \\repeatedly attempts to obtain an SSH account. \\Such attacks generate recurring authentication\\and connection activity. When attempts are \\distributed across multiple sources, each source \\may generate only a small number of failed login \\attempts, limiting the traffic volume attributable \\to any individual attacker~\cite{Sharafaldin2018TowardGA,javed2013detecting}.} \\
UNSW-NB15    & DoS           & {Malicious activity intended to make a server or \\network resource unavailable to legitimate users \\by interrupting or suspending its service \cite{mohammadian2024data}.~\\Flooding-based DoS attacks can achieve this by \\exhausting bandwidth, processing capacity, or \\server resources through sustained malicious \\traffic \cite{zargar2013survey}.}                                                              \\
             & Exploits      & {Attacks in which knowledge of a vulnerability in \\an operating system or software component \\is used to exploit the target. Because the category\\encompasses different exploitation behaviors, the \\resulting network activity can depend on the \\targeted service and vulnerability, and a uniform\\flow-level signature is not assumed \cite{mohammadian2024data}.}                                                \\
             & Recon         & {Information-gathering activity used to obtain \\knowledge about potential targets, including\\~network or service probing. UNSW-NB15\\represents reconnaissance as a broader attack \\category, and a uniform flow-level signature is not \\assumed \cite{mohammadian2024data}.}                                                                                                                                           
\end{tblr}
\end{table}

\section{Sensitivity Analysis}
\label{app:sensitivity}

Table \ref{tab:sensitivity} summarizes the sensitivity analysis of the DP-based detector to neighboring values of the selected thresholds.

\begin{table}[ht]
\centering
\small
\caption{Sensitivity of the DP detector to $\tau$ and $k$.
FPR is measured on benign traffic; Rec., BA, and MCC are averaged across
attack scenarios, while the last three columns report attack-specific recall.
Bold indicates the configuration used in the main evaluation.}
\label{tab:sensitivity}

\setlength{\tabcolsep}{2.5pt}
\begin{tabular}{ccrrrrrrr}
\toprule
\multicolumn{9}{c}{\textbf{CIC-IDS-2018}} \\
\midrule
$\tau$ & $k$ & FPR & Rec. & BA & MCC &
DDoS & Slow. & SSH BF \\
\midrule
2.0 & 1 & 0.49 & 0.98 & 0.75 & 0.44 & 0.94 & 1.00 & 1.00 \\
2.0 & 2 & 0.37 & 0.97 & 0.80 & 0.52 & 0.94 & 1.00 & 0.98 \\
2.0 & 3 & 0.30 & 0.81 & 0.76 & 0.40 & 0.94 & 1.00 & 0.50 \\
2.5 & 1 & 0.30 & 0.98 & 0.84 & 0.58 & 0.94 & 1.00 & 1.00 \\
\textbf{2.5} & \textbf{2} & \textbf{0.23} & \textbf{0.75} &
\textbf{0.76} & \textbf{0.44} & \textbf{0.94} & \textbf{0.80} &
\textbf{0.50} \\
2.5 & 3 & 0.17 & 0.58 & 0.71 & 0.32 & 0.44 & 0.79 & 0.50 \\
3.0 & 1 & 0.26 & 0.86 & 0.80 & 0.52 & 0.94 & 0.80 & 0.85 \\
3.0 & 2 & 0.17 & 0.69 & 0.76 & 0.44 & 0.94 & 0.79 & 0.34 \\
3.0 & 3 & 0.12 & 0.43 & 0.65 & 0.19 & 0.43 & 0.79 & 0.06 \\

\midrule
\multicolumn{9}{c}{\textbf{CIC-UNSW-NB15}} \\
\midrule
$\tau$ & $k$ & FPR & Rec. & BA & MCC &
DoS & Expl. & Recon. \\
\midrule
2.5 & 1 & 0.37 & 1.00 & 0.81 & 0.32 & 1.00 & 1.00 & 1.00 \\
2.5 & 2 & 0.24 & 1.00 & 0.88 & 0.41 & 0.99 & 1.00 & 1.00 \\
2.5 & 3 & 0.17 & 1.00 & 0.91 & 0.48 & 0.99 & 1.00 & 1.00 \\
3.0 & 1 & 0.30 & 1.00 & 0.85 & 0.37 & 1.00 & 1.00 & 1.00 \\
\textbf{3.0} & \textbf{2} & \textbf{0.18} & \textbf{1.00} &
\textbf{0.91} & \textbf{0.48} & \textbf{0.99} & \textbf{1.00} &
\textbf{1.00} \\
3.0 & 3 & 0.12 & 0.97 & 0.92 & 0.54 & 0.97 & 0.97 & 0.96 \\
3.5 & 1 & 0.28 & 1.00 & 0.86 & 0.38 & 1.00 & 1.00 & 1.00 \\
3.5 & 2 & 0.14 & 1.00 & 0.93 & 0.53 & 0.99 & 1.00 & 1.00 \\
3.5 & 3 & 0.06 & 0.85 & 0.89 & 0.59 & 0.86 & 0.88 & 0.80 \\
\bottomrule
\end{tabular}
\end{table}

\end{document}